\input harvmac
\noblackbox
\ifx\answ\bigans
\magnification=1200\baselineskip=14pt plus 2pt minus 1pt
\else\baselineskip=16pt 
\fi


\def\gs{g_{\rm string}}
\def\tb{type $IIB$\ }\def\ta{type $IIA$\ }\def\ti{type $I$\ }
\def\ap{\alpha'}

\def\ti{{\im\Tc^1}}
\def\tii{{\im\Tc^2}}
\def\tiii{{\im\Tc^3}}

\def\cf{{\it cf.\ }}
\def\ie{{\it i.e.\ }}
\def\eg{{\it e.g.\ }}
\def\eqq{{\it Eq.\ }}
\def\eqqs{{\it Eqs.\ }}
\def\th{\theta}

\def\Om{\Omega}
\def\om{\omega}

\def\Om{\Omega}

\newif\ifnref
\def\rrr#1#2{\relax\ifnref\nref#1{#2}\else\ref#1{#2}\fi}
\def\ldf#1#2{\begingroup\obeylines
\gdef#1{\rrr{#1}{#2}}\endgroup\unskip}
\def\nrf#1{\nreftrue{#1}\nreffalse}

\def\multrefv#1#2#3#4#5{\nrf{#1#2#3#4#5}\refs{#1{--}#5}}
\def\multrefiv#1#2#3#4{\nrf{#1#2#3#4}\refs{#1{--}#4}}

\def\doubref#1#2{\refs{{#1},{#2} }}
\def\threeref#1#2#3{\refs{{#1},{#2},{#3} }}
\def\fourref#1#2#3#4{\refs{{#1},{#2},{#3},{#4} }}

\nreffalse

\def\lref{\ldf}


\def\appA{A}

\def\tilde{\widetilde}

\def\h {{1\over 2}}

\def\ov {\overline}
\def\o {\over}
\def\fc#1#2{{#1 \o #2}}

\def\IZ{ {\bf Z}}
\def\IC{{\bf C}}

\def\hat{\widehat}

\def\br{\hfill\break}

\def\mod {{\rm mod}}
\def\lf {\left}
\def\ri {\right}
\def\ra {\rightarrow}

\def\im {{\rm Im}}
\def\p {\partial}

\def\Fc {{\cal F}} 
\def\Cc {{\cal C}} \def\Oc {{\cal O}}
 
\def\Mc {{\cal M}} 
 \def\Tc {{\cal T}}
 \def\Uc {{\cal U}}


\Title{\vbox{\rightline{MPP--2004--123}\rightline{LMU--TPS 04/10}
\rightline{\tt hep-th/0410074}}}
{\vbox{\centerline{MSSM with Soft SUSY Breaking Terms}
\bigskip\centerline{from $D7$--Branes with Fluxes}
}}
\smallskip
\centerline{D. L\"ust$^{a,b}$,\ \ S. Reffert$^{b}$, \ and\ \ S. Stieberger$^{a}$}
\bigskip
\centerline{\it $^a$ Department f\"ur Physik, Ludwig-Maximilians-Universit\"at M\"unchen,}
\centerline{\it Theresienstra{\ss}e 37, 80333 M\"unchen, FRG}
\vskip5pt
\centerline{\it $^b$ Max--Planck--Institut f\"ur Physik,}
\centerline{\it F\"ohringer Ring 6, 80805 M\"unchen, FRG}
\vskip15pt
\centerline{{\it E--mails:} {\tt luest,sreffert,stieberg@theorie.physik.uni-muenchen.de}}

\bigskip\bigskip
\centerline{\bf Abstract}
\vskip .2in
\noindent
We discuss the structure of the soft supersymmetry breaking terms
in a MSSM like model, which can be derived from
$D7$--branes with chiral matter fields from $2$--form $f$--fluxes 
and supersymmetry breaking from 3-form $G$--fluxes.

\Date{}
\noindent

\goodbreak

\lref\LustKS{
D.~L\"ust,
``Intersecting brane worlds: A path to the standard model?,''
Class.\ Quant.\ Grav.\  {\bf 21}, S1399 (2004)
[arXiv:hep-th/0401156].
}

\lref\LustFI{
  D.~Lust, S.~Reffert and S.~Stieberger,
  ``Flux-induced soft supersymmetry breaking in chiral type IIb orientifolds
  with D3/D7-branes,''
  Nucl.\ Phys.\ B {\bf 706}, 3 (2005)
  [arXiv:hep-th/0406092].
}

\lref\AFIV{G.~Aldazabal, A.~Font, L.E.~Ibanez and G.~Violero,
``D = 4, N = 1, type IIB orientifolds,''
Nucl.\ Phys.\ B {\bf 536}, 29 (1998)
[arXiv:hep-th/9804026].
}

\lref\CamaraJJ{
P.G.~Camara, L.E.~Ibanez and A.M.~Uranga,
``Flux-induced SUSY-breaking soft terms on D7-D3 brane systems,''
arXiv:hep-th/0408036.
}

\lref\IbanezIV{
L.E.~Ibanez,
``The Fluxed MSSM,''
arXiv:hep-ph/0408064.
}

\lref\FERRARA{R.~D'Auria, S.~Ferrara, F.~Gargiulo, M.~Trigiante and S.~Vaula,
``N = 4 supergravity Lagrangian for type IIB on $T^6/\IZ_2$ in presence of
fluxes and D3-branes,''
JHEP {\bf 0306}, 045 (2003)
[arXiv:hep-th/0303049];\br
R.~D'Auria, S.~Ferrara and S.~Vaula,
``N = 4 gauged supergravity and a IIB orientifold with fluxes,''
New J.\ Phys.\  {\bf 4}, 71 (2002)
[arXiv:hep-th/0206241].
}

\lref\ADFT{C.~Angelantonj, R.D'Auria, S.~Ferrara and M.~Trigiante,
``$K3 x T^2/\IZ_2$ orientifolds with fluxes, open string moduli and critical
points,''
Phys.\ Lett.\ B {\bf 583}, 331 (2004)
[arXiv:hep-th/0312019].
}

\lref\LMRS{
D. L\"ust, P.~Mayr, R.~Richter and S.~Stieberger,
``Scattering of gauge, matter, and moduli fields from intersecting branes,''
Nucl.\ Phys.\ B {\bf 696}, 205 (2004)
[arXiv:hep-th/0404134].
}

\lref\YUK{E.~Gava, K.S.~Narain and M.H.~Sarmadi,
``On the bound states of p- and (p+2)-branes,''
Nucl.\ Phys.\ B {\bf 504}, 214 (1997)
[arXiv:hep-th/9704006];\br
I.~Antoniadis, K.~Benakli and A.~Laugier,
"Contact interactions in $D$--brane models,''
JHEP {\bf 0105}, 044 (2001)
[arXiv:hep-th/0011281];\br
D.~Cremades, L.~E.~Ibanez and F.~Marchesano,
``Computing Yukawa couplings from magnetized extra dimensions,''
arXiv:hep-th/0404229;
``Yukawa couplings in intersecting D-brane models,''
JHEP {\bf 0307}, 038 (2003)
[arXiv:hep-th/0302105];\br
M.~Cvetic and I.~Papadimitriou,
"Conformal field theory couplings for intersecting $D$--branes on  orientifolds,''
Phys.\ Rev.\ D {\bf 68}, 046001 (2003)
[arXiv:hep-th/0303083];\br
I.R.~Klebanov and E.~Witten,
"Proton decay in intersecting $D$--brane models,''
Nucl.\ Phys.\ B {\bf 664}, 3 (2003)
[arXiv:hep-th/0304079]\br
S.A.~Abel and A.W.~Owen,
"Interactions in intersecting brane models,''
Nucl.\ Phys.\ B {\bf 663}, 197 (2003)
[arXiv:hep-th/0303124].
}

\lref\BCS{R.~Blumenhagen, J.P.~Conlon and K.~Suruliz,
``Type IIA orientifolds on general supersymmetric Z(N) orbifolds,''
JHEP {\bf 0407}, 022 (2004)
[arXiv:hep-th/0404254].
}

\lref\DijkstraYM{
T.P.T.~Dijkstra, L.R.~Huiszoon and A.N.~Schellekens,
 ``Chiral supersymmetric standard model spectra from orientifolds of Gepner
arXiv:hep-th/0403196.
}

\lref\HoneckerKB{
G.~Honecker and T.~Ott,
 ``Getting just the supersymmetric standard model at intersecting branes on the
Z(6)-orientifold,''
arXiv:hep-th/0404055.
}

\lref\GorlichQM{
L.~G\"orlich, S.~Kachru, P.K.~Tripathy and S.P.~Trivedi,
 ``Gaugino condensation and nonperturbative superpotentials in flux
arXiv:hep-th/0407130.
}

\lref\recently{A.~Font and L.E.~Ibanez,
``SUSY-breaking soft terms in a MSSM magnetized D7-brane model,''
  JHEP {\bf 0503}, 040 (2005)
  [arXiv:hep-th/0412150];\br
M.P.G.~del Moral,
``A new mechanism of Kahler moduli stabilization in type IIB theory,''
  arXiv:hep-th/0506116.
}

\lref\StiebergerYI{
S.~Stieberger,
``(0,2) heterotic gauge couplings and their M-theory origin,''
Nucl.\ Phys.\ B {\bf 541}, 109 (1999)
[arXiv:hep-th/9807124].
}

\lref\JockersYJ{
H.~Jockers and J.~Louis,
``The effective action of D7-branes in N = 1 Calabi-Yau orientifolds,''
arXiv:hep-th/0409098.
}

\lref\KachruSK{
S.~Kachru, M.B.~Schulz, P.K.~Tripathy and S.P.~Trivedi,
``New supersymmetric string compactifications,''
JHEP {\bf 0303}, 061 (2003)
[arXiv:hep-th/0211182].
}

\lref\BlumenhagenCG{
R.~Blumenhagen and T.~Weigand,
``Chiral supersymmetric Gepner model orientifolds,''
JHEP {\bf 0402}, 041 (2004)
[arXiv:hep-th/0401148].
}

\lref\AngelantonjHI{
C.~Angelantonj, I.~Antoniadis, E.~Dudas and A.~Sagnotti,
``Type-I strings on magnetised orbifolds and brane transmutation,''
Phys.\ Lett.\ B {\bf 489}, 223 (2000)
[arXiv:hep-th/0007090];\br
C.~Angelantonj and A.~Sagnotti,
``Type-I vacua and brane transmutation,''
arXiv:hep-th/0010279.
}

\lref\BachasIK{
C.~Bachas,
``A Way to break supersymmetry,''
arXiv:hep-th/9503030.
}

\lref\zwart{G.~Zwart,
``Four-dimensional N = 1 Z(N) x Z(M) orientifolds,''
  Nucl.\ Phys.\ B {\bf 526}, 378 (1998)
  [arXiv:hep-th/9708040].
}

\lref\GVW{S.~Gukov, C.~Vafa and E.~Witten,
``CFT's from Calabi-Yau four-folds,''
Nucl.\ Phys.\ B {\bf 584}, 69 (2000)
[Erratum-ibid.\ B {\bf 608}, 477 (2001)]
[arXiv:hep-th/9906070].
}

\lref\PM{P.~Mayr,
``Stringy world branes and exponential hierarchies,''
JHEP {\bf 0011}, 013 (2000)
[arXiv:hep-th/0006204].
}

\lref\wip{Work in progress.}

\lref\ABFPT{I.~Antoniadis, C.~Bachas, C.~Fabre, H.~Partouche and T.R.~Taylor,
``Aspects of type I - type II - heterotic triality in four dimensions,''
Nucl.\ Phys.\ B {\bf 489}, 160 (1997)
[arXiv:hep-th/9608012].
}

\lref\CIM{ D.~Cremades, L.E.~Ibanez and F.~Marchesano,
``More about the standard model at intersecting branes,''
  arXiv:hep-ph/0212048;
``Yukawa couplings in intersecting D-brane models,''
  JHEP {\bf 0307}, 038 (2003)
  [arXiv:hep-th/0302105].
}

\lref\LRSm{D.~Lust, P.~Mayr, S.~Reffert and S.~Stieberger,
  ``F-theory flux, destabilization of orientifolds and soft terms on
  D7-branes,''
  arXiv:hep-th/0501139.
}

\lref\LRS{D.~Lust, S.~Reffert, W.~Schulgin and S.~Stieberger,
``Moduli stabilization in type IIB orientifolds. I: Orbifold limits,''
  arXiv:hep-th/0506090.
}

\lref\MayrHH{
P.~Mayr,
 ``On supersymmetry breaking in string theory and its realization in brane
worlds,''
Nucl.\ Phys.\ B {\bf 593}, 99 (2001)
[arXiv:hep-th/0003198].
}

\lref\TV{T.R.~Taylor and C.~Vafa,
``RR flux on Calabi-Yau and partial supersymmetry breaking,''
Phys.\ Lett.\ B {\bf 474}, 130 (2000)
[arXiv:hep-th/9912152].
}

\lref\KorsWF{
B.~K\"ors and P.~Nath,
``Effective action and soft supersymmetry breaking for intersecting D-brane
Nucl.\ Phys.\ B {\bf 681}, 77 (2004)
[arXiv:hep-th/0309167].
}

\lref\CurioSC{
G.~Curio, A.~Klemm, D.~L\"ust and S.~Theisen,
``On the vacuum structure of type II string compactifications on  Calabi-Yau
spaces with H-fluxes,''
Nucl.\ Phys.\ B {\bf 609}, 3 (2001)
[arXiv:hep-th/0012213].
}

\lref\AldazabalDG{
G.~Aldazabal, S.~Franco, L.E.~Ibanez, R.~Rabadan and A.M.~Uranga,
``D = 4 chiral string compactifications from intersecting branes,''
J.\ Math.\ Phys.\  {\bf 42}, 3103 (2001)
[arXiv:hep-th/0011073].
}

\lref\MaSh{
F. Marchesano and G. Shiu,
``MSSM vacua from flux compactifications,''
arXiv:hep-th/0408059;
``Building MSSM flux vacua,''
arXiv:hep-th/0409132.
}

\lref\BL{M.~Berkooz and R.G.~Leigh,
``A D = 4 N = 1 orbifold of type I strings,''
Nucl.\ Phys.\ B {\bf 483}, 187 (1997)
[arXiv:hep-th/9605049].
}

\lref\BDL{M.~Berkooz, M.R.~Douglas and R.G.~Leigh,
"Branes intersecting at angles,''
Nucl.\ Phys.\ B {\bf 480}, 265 (1996)
[arXiv:hep-th/9606139].
}
\lref\GKP{S.B.~Giddings, S.~Kachru and J.~Polchinski,
``Hierarchies from fluxes in string compactifications,''
Phys.\ Rev.\ D {\bf 66}, 106006 (2002)
[arXiv:hep-th/0105097].
}

\lref\FP{A.R.~Frey and J.~Polchinski,
``N = 3 warped compactifications,''
Phys.\ Rev.\ D {\bf 65}, 126009 (2002)
[arXiv:hep-th/0201029].
}

\lref\uranga{M.~Cvetic, G.~Shiu and A.~M.~Uranga,
``Chiral four-dimensional N = 1 supersymmetric type IIA orientifolds from
intersecting D6-branes,''
Nucl.\ Phys.\ B {\bf 615}, 3 (2001)
[arXiv:hep-th/0107166].
}

\lref\GL{T.W.~Grimm and J.~Louis,
``The effective action of N = 1 Calabi-Yau orientifolds,''
arXiv:hep-th/0403067.
}

\lref\JGL{M.~Grana, T.W.~Grimm, H.~Jockers and J.~Louis,
``Soft supersymmetry breaking in Calabi-Yau orientifolds with D-branes and
fluxes,''
arXiv:hep-th/0312232.
}

\lref\BlumenhagenWH{
R.~Blumenhagen, L.~G\"orlich, B.~K\"ors and D.~L\"ust,
``Noncommutative compactifications of type I strings on tori with  magnetic
background flux,''
JHEP {\bf 0010}, 006 (2000)
[arXiv:hep-th/0007024].
}

\lref\CveticXX{
M.~Cvetic and T.~Liu,
``Three-family supersymmetric standard models, flux compactification and moduli
stabilization,''
arXiv:hep-th/0409032.
}

\lref\CIU{P.G.~Camara, L.E.~Ibanez and A.M.~Uranga,
``Flux-induced SUSY-breaking soft terms,''
arXiv:hep-th/0311241.
}

\lref\KST{S.~Kachru, M.B.~Schulz and S.~Trivedi,
``Moduli stabilization from fluxes in a simple IIB orientifold,''
JHEP {\bf 0310}, 007 (2003)
[arXiv:hep-th/0201028].
}

\lref\CremmerEN{
E.~Cremmer, S.~Ferrara, L.~Girardello and A.~Van Proeyen,
 ``Yang-Mills Theories With Local Supersymmetry: Lagrangian, Transformation
Nucl.\ Phys.\ B {\bf 212}, 413 (1983).
}

\lref\LS{D.~L\"ust and S.~Stieberger,
``Gauge threshold corrections in intersecting brane world models,''
arXiv:hep-th/0302221.
}

\lref\BLT{R.~Blumenhagen, D.~L\"ust and T.R.~Taylor,
``Moduli stabilization in chiral type IIB orientifold models with fluxes,''
Nucl.\ Phys.\ B {\bf 663}, 319 (2003)
[arXiv:hep-th/0303016].
}

\lref\PolchinskiSM{
J.~Polchinski and A.~Strominger,
``New Vacua for Type II String Theory,''
Phys.\ Lett.\ B {\bf 388}, 736 (1996)
[arXiv:hep-th/9510227].
}

\lref\MichelsonPN{
J.~Michelson,
Nucl.\ Phys.\ B {\bf 495}, 127 (1997)
[arXiv:hep-th/9610151].
}

\lref\CU{J.F.G.~Cascales and A.M.~Uranga,
``Chiral 4d N = 1 string vacua with D-branes and NSNS and RR fluxes,''
JHEP {\bf 0305}, 011 (2003)
[arXiv:hep-th/0303024].
}

\lref\ADFT{C.~Angelantonj, R.~D'Auria, S.~Ferrara and M.~Trigiante,
``$K_3 \times T^2/\IZ_2$ orientifolds with fluxes, open string moduli and critical
points,''
Phys.\ Lett.\ B {\bf 583}, 331 (2004)
[arXiv:hep-th/0312019].
}

\lref\SOFTref{
L.E.~Ibanez and D.~L\"ust,
"Duality anomaly cancellation, minimal string unification and the effective low-energy Lagrangian 
of 4-D strings,''
Nucl.\ Phys.\ B {\bf 382}, 305 (1992)
[arXiv:hep-th/9202046];\br
V.S.~Kaplunovsky and J.~Louis,
"Model independent analysis of soft terms in effective supergravity and in string theory,''
Phys.\ Lett.\ B {\bf 306}, 269 (1993)
[arXiv:hep-th/9303040].
}
\lref\BIM{A.~Brignole, L.E.~Ibanez and C.~Munoz,
"Towards a theory of soft terms for the supersymmetric Standard Model,''
Nucl.\ Phys.\ B {\bf 422}, 125 (1994)
[Erratum-ibid.\ B {\bf 436}, 747 (1995)]
[arXiv:hep-ph/9308271].
}

\newsec{Introduction}

Whether the minimal supersymmetric standard model
(MSSM) or some of its ramifications will be experimentally discovered at the
LHC is of burning 
interest also for theoretical particle physics. In the MSSM, supersymmetry
breaking
is usually parametrized by a set of soft SUSY breaking parameters,
like gaugino, squark and slepton masses, 
which have the virtue that they
do not spoil the good renormalization behaviour of supersymmetric
field theories. But the MSSM does not offer any deeper microscopic explanation
of the origin of the soft SUSY breaking parameters.
Nevertheless there are some phenomenological constraints on
the structure of the soft terms, e.g. the absence of flavor
changing neutral currents strongly favors squark masses, 
which are universal for all squark flavors.

As is well known, 
a controllable way to obtain the soft supersymmtry breaking terms of the MSSM
is provided by coupling the matter sector of the MSSM to local
N=1 supergravity. Then 
spontaneous supersymmetry breaking  
by non-vanishing $F$-- or $D$--terms induces soft supersymmetry breaking terms in the
matter field action.  
Superstring theory offers a concrete, microscopic realization 
of soft SUSY breaking in N=1 supergravity:
the effective low energy action of supersymmetric string compactifications
to four space-time dimensions is given by the 
N=1 supergravity action of \CremmerEN.
Furthermore, spontaneous supersymmetry breaking is due to $F$--terms
of the gauge singlet scalar fields,
namely the
dilaton $S$ or the geometric moduli $M$, whose $F$--terms
are called $F^S$ and $F^{M}$, respectively.
 Then supersymmetry breaking is transmitted
from the gauge neutral sector to the charged sector of the MSSM
by gravitational interactions. This scenario already allows
for a fairly model independent analysis of the soft terms, which are all
proportional to certain combinations of  $F^S$ or $F^{M}$ 
\SOFTref. In particular, the dilaton dominated scenario
with $F^S\neq 0$, $F^{M}=0$ possesses the feature
of flavor universal soft scalar masses, which is usually spoilt
by non-vanishing vevs for $F^{M}$.
In more generic
scenarios, in which both  $F^S\neq 0$ and $F^{M}\neq 0$, the
soft SUSY breaking terms
can be nicely parametrized
by a so-called goldstino angle $\tan\theta_g\sim F^S/F^T$ \BIM, where
$T$ is the overall volume modulus of the internal space.

The final step for a complete understanding of the soft-terms
is undertaken by knowing 
$(i)$ how the matter sector of the MSSM is microscopically built in string
theory, and $(ii)$ how the supersymmetry breaking auxiliary fields $F^S$,
$F^{M}$ 
are induced, i.e. by knowing how a non-trivial effective superpotential
for the fields $S$ and $M$ is generated.
In this paper we are interested in compactifications
of the type $I$ strings, namely the so-called orientifold compactifications of
the type $IIA/B$ superstring.
Let us first recall how point $(i)$ 
can be realized in orientifold compactifications. Namely 
one very promising way to build the MSSM is to use
intersecting $D6$--branes in \ta orientifolds (for a  review see \LustKS).
The gauge degrees of freedom are due to open strings living
on each of the various stacks of $D6$--branes, whereas the chiral matter fields
are localized on the lower-dimensional intersection loci of the
$D6$--branes. More specifically, the $D6$--branes, all completely filling 
four-dimensional Minkowski space-time, are wrapped around supersymmetric
3-cycles in the internal space $X_6$, which generically intersect just
on points in $X_6$. Note that the internal intersection numbers are
normally larger than one, a fact, which offers a nice explantion for
the family replication of the MSSM.
In order to preserve N=1 space-time supersymmetry 
in the open string sectors on the
intersecting $D$--branes, the intersection
angles must obey certain conditions, and for consistent model building, all
Ramond tadpoles must be cancelled.
Starting from the original work on non-supersymmetric
models   \multrefv\BDL\BachasIK\BlumenhagenWH\AngelantonjHI\AldazabalDG, 
several semirealistic MSSM-like models with intersecting
$D6$--branes
were constructed during the last years
\multrefiv\uranga\DijkstraYM\HoneckerKB\BlumenhagenCG.
However, for practical reasons, when turning on the SUSY-breaking 3-form fluxes
(see later), it is more convenient to use instead of the \ta
orientifolds with intersecting $D6$--branes the mirror ($T$--dual) \tb
orientifold description. Then, after an appropriate mirror
transformation, the (supersymmetric) $D6$--branes are transformed into a system of
$D3$--branes plus supersymmetric $D7$--branes, where the non-trivial intersection
angles
in \ta become open string 2-form gauge fluxes (magnetic $f$-field
background fields) living on internal 4-cycles on the different
$D7$--brane world volumes. 
Note that $f$--fluxes are required at least on some
of the various stacks of $D7$--branes in order
to obtain realistic models with more than one {\sl chiral} generation
of quarks and leptons. Hence for getting chiral fermions, some of the $D7$--branes
possess mixed Dirichlet/Neumann boundary conditions in certain
internal directions, which means that they are a kind of
hybrid between $D3$-- and $D7$--branes.
This fact will become important for the structure of
the soft terms for the matter fields on the $D3/D7$--brane world volumes.
  
Now coming to the second issue $(ii)$ of spontaneous supersymmetry
breaking we will consider the generation of an effective superpotential 
\multrefiv\GVW\TV\MayrHH\CurioSC\
for the dilaton $S$ and the moduli fields $M$ by flux compactifications
\doubref\PolchinskiSM\MichelsonPN\
with non-vanishing, internal fluxes
of the \tb 3-form $G_3=F_3-SH_3$, where $F_3$ and $H_3$ are the field strengths
of the Ramond and the Neveau--Schwarz 2-form gauge potentials $C_2^{R}$, $B_2^{NS}$, respectively.
As it was shown in \doubref\GKP\KST\ ,
the 3-form fluxes 
in general contribute to
the tadpole conditions, but still preserve N=1 supersymmetry, \ie
$F^S=F^{M}=0$ in the vacuum, if 
$G_3$ is imaginay self-dual ($ISD$ flux) and of Hodge type $(2,1)$ on the internal Calabi-Yau 
space. However all complex structure moduli $U^i$ as well as the dilaton
field $S$ are already fixed in a generic supersymmetric flux vacuum.
On the other hand, if $G_3$ is an $ISD$ $(0,3)$-form, it
corresponds to a non-vanishing auxiliary field $F^T$ of the overall K\"ahler modulus
of $X_6$, and
supersymmetry is spontaneously broken;
if $G_3$ is an imaginary anti-self dual ($IASD$) $(3,0)$-form
it is equivalent to an auxiliary field $F^S$; finally if $G_3$ contains some of the $IASD$ $(1,2)$ 
forms, this is described by non-vanishing auxiliary fields $F^{U^i}$ for
the complex structure moduli $U^i$ in the effective field theory
description.
In the following, we will mainly concentrate on the two cases of
$ISD$ $(0,3)$--flux and/or $IASD$ $(3,0)$--fluxes, since these are the generic 3-form fluxes
for all $\IZ_N\times \IZ_M$ orientifold compactifications.

In order to derive the soft SUSY breaking parameters, one has
to compute the couplings between the open string matter fields on the
$D3/D7$--branes and the closed string
$3$--form field strengths $G_3$. 
Explicit \tb orientifold models
with $ISD$-fluxes have been already constructed in \fourref\KST\KachruSK\GorlichQM\FERRARA, 
with chirality in \threeref\BLT\CU\LustFI.  
Recently, MSSM--like flux models with $D3/D7$--branes (including magnetized $D9$--branes) 
and complete cancellation of both $R$-- and $NS$--tadpoles
have been constructed in {\it Ref.} \MaSh\ (see also {\it Ref.} \CveticXX).
The four-dimensional N=1 effective action of orientifolds
with $D3$- and/or $D7$-branes can be obtained by
the calculation of the
open/closed string scattering
amplitudes \LMRS\
or by Kaluza-Klein reduction of
the Dirac-Born-Infeld and Chern-Simons action
\doubref\GL\JockersYJ.
Then  the soft SUSY breaking terms can be derived
either by studying the Born-Infeld action
on the $D$--brane world volumes coupled to the flux $G_3$ as accomplished 
for $D3$--branes in \doubref\CIU\JGL, and for $D7$--branes in 
\doubref\CamaraJJ\IbanezIV, or by coupling the effective action
from open/closed string scattering
amplitudes to the effective closed
string action with $3$--form fluxes turned on, as it was performed 
for $D3$-- and $D7$--branes in \LustFI\ (see also 
the discussion of soft terms
in intersecting brane word models in {\it Ref.} \KorsWF).
Note that  in  {\it Ref.}  \LustFI\  also the open string
2--form $f$--flux on the $D7$--branes has been taken into account, which is crucial
for realistic model building with chiral fermions.
In any case, the results of the two different approaches \LustFI\ and \doubref\CamaraJJ\IbanezIV\
are completely consistent with each other 
and lead to identical results for vanishing $f$--flux.
The results can be summarized as follows:

$\bullet$ {\sl gaugino masses:}
Since for $D3$--branes the gauge kinetic function is given as $f\sim S$,
the $D3$--brane gaugino masses are sensitive to non-vanishing $IASD$ $(3,0)$-flux
with $F^S\neq 0$, but still vanish for non-trivial $ISD$ $(0,3)$-flux.
This situation is reversed for $D7$--branes with zero $f$--flux: their gauge
kinetic function is proportial to the transversal K\"ahler moduli $T^i$,
$f_i\sim T^i$, and hence the gaugino masses only feel the $ISD$ $(0,3)$-flux
with $F^T\neq 0$. So the role of $D3$--branes and pure $D7$--branes is reversed.
Note however that for $D7$--branes with non-vanishing $f$--flux, i.e.
with mixed $D/N$--boundary conditions, the gauge kinetic function
contains both the dilaton $S$ as well as the
K\"ahler moduli $T^i$. Therefore the corresponding
gaugino masses get contributions both from the $(0,3)$ and also
from the $(3,0)$-flux, as it will happen in realistic models with
three chiral generations. Hence, for the $D7$--branes with mixed boundary conditions
it will be convenient to parametrize the gaugino masses
by the goldstino angle $\sin\theta_g$.

$\bullet$ {\sl scalar masses:} the soft scalar masses follow a similar pattern
as compared to the gaugino masses.
For the scalars living on the $D3$--branes, a mass is only generated by the
(3,0)-flux, while scalars on pure $D7$--branes get their masses partly also 
from $(0,3)$-flux. On the other hand, scalars on $D7$--branes with
$f$--fluxes get mass contributions both from $(3,0)$- and$(0,3)$-fluxes.
Most importantly, `chiral' scalar fields, which correspond to twisted 
open string sectors, \ie open strings which stretch 
between two $D7$--branes with
different type of $f$--flux boundary conditions, get also masses from $(3,0)$- 
as well as from$(0,3)$-fluxes.


The outline of our work is the following:
In the next section we shall recall the general structure of the
3-form flux induced soft terms for $D7$--branes  with $f$--flux, following our
previous work in {\it Ref.} \LustFI. 
In section three, we shall make an attempt to gather some
generic information on the structure of the
soft terms in MSSM--like orientifold constructions. Here
our strategy is the following.
Instead of considering compact models which satisfy all tadpole
conditions,
we will rather consider a locally supersymmetric $D7$--brane
set-up, which precisely contains the open string matter
fields with three generation MSSM quantum numbers.  Specifically, a minimal
way to build the MSSM via three stacks of
intersecting $D6$--branes on a six-torus $T^6$
(or also on an orbifold) was proposed in
\CIM. We will use this type $IIA$ setup, perform
the mirror transformation to \tb and will derive
the equivalent brane configuration. The latter now consists
of three different stacks of $D7$--branes, one being equipped
with non-trivial open string 2-form $f$--flux.
Via this rather simple construction we can finally compute
all relevant soft terms.
In section 4 we shall parametrize our results by the goldstino angle, while
in section 5 we discuss the scales of the gravitino mass and soft--masses induced by 
non--vanishing $(0,3)$ and $(3,0)$--form fluxes.
Finally, in section 6 we give some concluding remarks.

\newsec{Soft terms for $D7$--branes  with $f$--flux}

In this chapter, we will recall the general structure of the soft terms
for matter fields originating from  $D3$-- or $D7$--branes with $f$--fluxes.
We will follow the approach of {\it Ref.} \LustFI, where these terms
have been determined by computing the effective action for
the open string matter fields by a direct calculation
of string scattering amplitudes, and subsequently coupling
the matter fields to the 3--form flux induced superpotential. An alternative
derivation of the soft terms for $D7$--branes without $f$--flux
using the Born--Infeld action can be found in \CamaraJJ.

\subsec{Three--form $G$--flux}

Let us start by reviewing the main aspects of \tb 3-form fluxes and
the corresponding superpotential. We concentrate on orientifolds of \tb compactified on the
toroidal orbifold
\eqn\orbi{
X_6=\fc{T^6}{\IZ_N\times \IZ_M}\ ,}
with the orbifold group $\Gamma=\IZ_N\times \IZ_M$. 
There are $h_{(1,1)}(X_6)$ K\"ahler moduli and $h_{(2,1)}(X_6)$ complex
structure moduli, which split into twisted and untwisted moduli. In the following 
the dimension of the latter is denoted by $h^{untw.}_{(1,1)}(X_6)$ and 
$h^{untw.}_{(2,1)}(X_6)$, respectively.
In addition, there is the complex dilaton field $S$.
To obtain an N=1 (closed) string spectrum, one introduces an 
orientifold projection $\Om I_n$, with $\Om$ describing a reversal of the 
orientation of the closed string world--sheet and $I_n$ a reflection of $n$ internal
coordinates. For $\Om I_n$ to represent a symmetry of the original theory, $n$
has to be an even integer in \tb.
Generically, this projection produces orientifold fixed planes [$O(9-n)$--planes],
placed at the orbifold fixpoints of $T^6/I_n$. They have negative tension, which 
has to be balanced by introducing positive tension objects.
Candidates for the latter may be collections of $D(9-n)$--branes and/or non--vanishing
three--form fluxes $H_3$ and $C_3$.
The orbifold group $\Gamma$ mixes with the orientifold group $\Om I_n$.
As a result, if the group $\Gamma$ contains $\IZ_2$--elements $\th$, 
which leave one complex plane fixed, we obtain additional $O(9-|n-4|)$-- or 
$O(3+|n-2|)$--planes from the element $\Om I_n \th$. 

In the following, only the two cases of $n=6$ ($O3$--plane) and $n=2$
($O7$--planes) will be relevant
to us. Tadpoles may be completley cancelled by adding $D3/D7$--branes, 
provided the orbifold twist $\Gamma$ is $\IZ_2\times \IZ_2,\IZ_2\times\IZ_3,\IZ_2\times \IZ_6,
\IZ_2\times\IZ_6',\IZ_3,\IZ_3\times
\IZ_3,\IZ_6-I,\IZ_6-II,\IZ_3\times \IZ_6,\IZ_6\times \IZ_6,\IZ_7$ or $\IZ_{12}-I$ 
\doubref\zwart\AFIV. 
This is to be contrasted with \ta intersecting $D6$--brane
constructions, where it has been recently shown that essentially all orbifold
groups $\Gamma$ allow for tadpole cancellation due to the appearance of only
untwisted and $\IZ_2$--twisted sector tadpoles \BCS.

Let us now give non--vanishing vevs to some of the (untwisted) 
flux components $H_{ijk}$ and $F_{ijk}$, with $F_3=dC_2$, $H_3=dB_2$. 
The two $3$--forms $F_3,H_3$ are organized in the $SL(2,\IZ)_S$ covariant field:
\eqn\fluxcomb{
G_3=F_3-S H_3\ .}
On the torus $T^6$, we would have 20+20 independent internal
components for $H_{ijk}$ and $F_{ijk}$. 
However, only a portion of them is invariant under the orbifold group $\Gamma$.
More precisely, of the $20$ complex (untwisted) components comprising the flux $G_3$, only 
$2h^{untw.}_{(2,1)}(X_6)+2$ survive the orbifold twist. 
The orientifold action $\Om(-1)^{F_L}I_6$ producing $O3$--planes does not give rise to any 
further restrictions. If the orbifold group $\Gamma$ contains 
$\IZ_2$--elements $\th$, which leave the $j$--th complex plane fixed, we also 
encounter $O7_j$--planes transverse to the $j$--th plane. Since $I_2^j=I_6\th$, the 
orientifold generator $\Om(-1)^{F_L}I_2^j$ does not put further restrictions on the 
$2h^{untw.}_{(2,1)}(X_6)+2$ twist invariant  components.
Hence, the allowed flux components are most conveniently found in the complex basis, in which
the orbifold group $\Gamma$ acts diagonally. In the following we shall
concentrate\foot{As an example, we may take the orientifold with 
orbifold group $\Gamma=\IZ_2\times \IZ_2$, discussed in {\it Ref.} \LustFI.
For this compactification we have $h^{untw.}_{(2,1)}(X_6)=3$. Hence we have 
$8+8$ untwisted flux components $H_{ijk}$ and $F_{ijk}$.} 
on the \tb orientifold/orbifolds $T^6/(\Gamma+\Gamma\Om I_6)$, with $\Gamma$ being
one of the (consistent) orbifold twists encountered above. Note, that $O7$--planes appear, 
in the case, that the orbifold twist $\Gamma$ is of even order.

The most general (untwisted) $3$--form flux $G_3$ 
may be written as linear combination of the complex cohomology 
group  $H^3(X_6,\IC)$ 
\eqn\GC{{1\over{(2\pi)^2\alpha'}}\ G_3=\sum_{i=0}^3 (A^i\omega_{Ai}+B^i\omega_{Bi})\ ,}
with a basis of $H^3=H^{(3,0)}\oplus H^{(2,1)}\oplus H^{(1,2)}\oplus H^{(0,3)}$:
\eqn\cplxz{\eqalign{
\om_{A_0}&=d\ov z^1\wedge d\ov z^2\wedge d\ov z^3\ \ \ ,\ \ \ \om_{A_1}=d\ov z^1\wedge
dz^2\wedge dz^3\ ,\cr
 \om_{A_2}&=dz^1\wedge
d\ov z^2\wedge dz^3\ \ \ ,\ \ \ \om_{A_3}=dz^1\wedge dz^2\wedge d\ov z^3\ ,\cr
\om_{B_0}&=dz^1\wedge d z^2\wedge d z^3\ \ \ ,\ \ \ 
\om_{B_1}= dz^1\wedge d\ov z^2\wedge d\ov z^3\ ,\cr
 \om_{B_2}&=d\ov z^1\wedge dz^2\wedge d\ov z^3\ \ \ ,\ \ \ 
\om_{B_3}= d\ov z^1\wedge d\ov z^2\wedge dz^3\ .}}
The expansion \GC\ is to be understood such, that according to the discussion from above only the
twist invariant $3$--forms contribute in the sum.
In the form \cplxz\ the cohomology structure of $G_3$ is manifest, but the
$SL(2,\IZ)_S$--covariance is not.
The form $\om_{A_0}$ corresponds to the $(0,3)$--part of the
flux, the $\om_{A_i},\ i=1,2,3,$ correspond to the $(2,1)$--part, $\om_{B_0}$ comprises
the $(3,0)$--part and the $\om_{B_i}\ i=1,2,3$  the $(3,0)$--part.
All twist--invariant fluxes fulfill the primitivity condition $G_3\wedge J=0$, with
$J$ the K\"ahler form (for more details we refer to {\it Ref.} \LRS).

In order to impose flux quantization on $G_3$, one has to transform the forms \cplxz\
into a real basis of 3--forms $H^3(T^6,\IZ)$ on $T^6$:
\eqn\realbase{
\eqalign{\alpha_0&=dx^1 \wedge dx^2  \wedge dx^3\ \ \ ,\ \ \ 
\beta^0=dy^1 \wedge dy^2 \wedge dy^3\ ,\cr
\alpha_1&=dy^1 \wedge dx^2  \wedge dx^3\ \ \ ,\ \ \ \beta^1=-dx^1 \wedge dy^2\wedge dy^3\ ,\cr
\alpha_2&=dx^1 \wedge dy^2  \wedge dx^3\ \ \ ,\ \ \ \beta^2=-dy^1 \wedge dx^2 \wedge dy^3\ ,\cr
\alpha_3&=dx^1 \wedge dx^2  \wedge dy^3\ \ \ ,\ \ \ \beta^3=-dy^1 \wedge dy^2 \wedge dx^3\ .}}
with the six real periodic coordinates $x^i,y^i$ on the torus $T^6$, \ie 
$x^i\cong x^i+1$ and $y^i\cong y^i+1$.
This is achieved through introducing complex structures:
\eqn\achieved{
dz^j=\sum_{i=1}^3\ \rho^j_i\ dx^i+ \tau^j_i\ dy^i\ \ \ ,\ \ \ j=1,2,3\ .}
Most of the parameters $\rho^j_i$ and $\tau^j_i$ are fixed through the orbifold twist $\Gamma$,
with  only those remaining undetermined, which correspond to the $\IZ_2$--elements of $\Gamma$.
The latter are eventually fixed through the flux quantization condition.
For further details see {\it Ref.} \LRS.
The basis \realbase\ has the property $\int_{X_6} \alpha_i \wedge \beta^j=\delta^j_i$.
Expressed in this basis,  the $G_3$-flux \fluxcomb\ takes the following form:
\eqn\GR{{1\over{(2\pi)^2\alpha'}}{G_3}=\sum_{i=0}^{3} \lf[(a^i-S
c^i)\alpha_i+(b_i-S d_i)\beta^i\ri]\ .}
In this basis, the  $SL(2,\IZ)_S$--covariance of $G_3$ is manifest. 
The coefficients $a^i,\ b_i$ refer to the Ramond part of $G_3$, whereas the
coefficients $c^i,\ d_i$ refer to the Neveu-Schwarz part.

As described above, not all of the eight flux components in \GR\ or \GC\ survive
the orbifold projection. In addition, some or all complex structure moduli
are frozen to discrete values by the $\IZ_N\times \IZ_M$ modding (see {\it Ref.} \LRS\ for more 
details).
On the other hand, in the $\IZ_2\times \IZ_2$ orbifold
all eight flux components survive and all three complex structure moduli $U^j\ ,\ j=1,2,3$ 
remain unfixed. 
However \eg in the $\IZ_3$ orbifold, only the components 
$G_{(3,0)}$ and $G_{(0,3)}$ are allowed, and all $U^i$ are frozen to
$U^i=\rho:=\half +\fc{i}{2}\sqrt 3$. 
Only the $IASD$--flux $G_{(3,0)}$ and the $ISD$--flux $G_{(0,3)}$
are generic flux components being invariant under all possible orbifold groups \LRS.
Let us remark, that due to the absence of the $ISD$ $(2,1)$ 3--form fluxes in most of
the $\IZ_N$--orbifold models, supersymmetric flux solutions do not exist for these cases.
Therefore, we shall concentrate in the following discussion on these two complex
fluxes, which are parametrized by four real coefficients.
Expressed in terms of the complex basis \GC,
the $G_{(3,0)}$ and $G_{(0,3)}$ fluxes take the following form:
\eqn\eqnulldreicplx{\eqalign{
{1\over{(2\pi)^2\alpha'}}\ G_{03}&=A_0\ \om_{A0}=A_0\ (d\ov z^1\wedge d\ov z^2\wedge d\ov z^3)
\, ,\cr
{1\over{(2\pi)^2\alpha'}}\ G_{30}&=B_0\ \om_{B0}=B_0\ (dz^1\wedge
dz^2\wedge dz^3)\, .}}

\subsec{Closed string low--energy effective action}

Now we shall consider the low-energy effective action of the 
closed string moduli fields $M$ for non--vanishing $G_{(3,0)}$-- or $G_{(0,3)}$--flux,
bearing in mind that some or all of the complex structure moduli may be frozen
to specific values in many of the orbifold compactifications.
Here, $M$ collectively accounts for the closed string moduli fields $S,\ T^j,\ U^j$.
The kinetic energy terms of these bulk fields are derived from the K\"ahler potential
$\hat K$ given by 
\doubref\ABFPT\LMRS:
\eqn\Khut{
\kappa_4^2\hat{K}(M,\ov M)=-\ln(S-\ov{S})-\sum_{j=1}^{h_{(1,1)}^{untw.}}
\ln(T^j-\ov{T}^j)-\sum_{j=1}^{h_{(2,1)}^{untw.}}\ln(U^j-\ov{U}^j)\ .}
The moduli fields $M$ refer to complex scalars of N=1 chiral multiplets. These fields $M$
have a functional dependence on the moduli fields $\Mc$ one uses in string--theory.
The latter, which will be introduced in {\it Ref.} \LRS\ through their geometric meaning
refer to the vertex operators following from the $\sigma$--model and are used
to study duality symmetries. For toroidal \tb orientifolds with $D3$-- and $D7$--branes
and $h_{(1,1)}^{untw.}=3$ we have the following relations
\eqn\fieldST{\eqalign{
T^j&=a^j+i\,{e^{-\phi_4}\over {2\pi\alpha'^{1/2}}}
\sqrt{{\im\,\Tc^k\im\,\Tc^l\over{\im\,\Tc^j}}}\ ,\cr
S&=C_0+i\,{e^{-\phi_4}\over {2\pi}}{\alpha'^{3/2}
\over{\sqrt{\im\,\Tc^1\im\,\Tc^2\im\,\Tc^3}}}\ ,\cr
U^j&=\Uc^j\ \ \ ,\ \ \ j=1,2,3\ ,}}
with the geometric (untwisted) K\"ahler moduli $\Tc^j$ and complex structure moduli $\Uc^j$
to be specified in \LRS. Here the axion follows from integrating the Ramond $4$--form 
over a $4$--cycle: $a^j=\int\limits_{T^{2,k}\times T^{2,l}} C_4$.
The K\"ahler potential $\Khut$ is quite generic for 
all type $II$ toroidal orbifolds, which essentially
only differ\foot{Note, that $h_{(1,1)}^{untw.}=3$
for almost all $\IZ_N\times \IZ_M$--orbifolds, except: $h_{(1,1)}^{untw.}(T^6/\IZ_3)=9$ and 
$h_{(1,1)}^{untw.}(T^6/\IZ_6-I)=5$. In these two special cases
the K\"ahler potential \Khut\ describes only the three diagonal K\"ahler moduli.
Furthermore, $h_{(2,1)}^{untw.}\leq 1$, except $h_{(2,1)}^{untw.}(T^6/\IZ_2\times \IZ_2)=~3$. For
further information see {\it Ref.} \LRS.} 
by their number $h_{(1,1)}^{untw.}$ of K\"ahler $T^j$ and their 
number $h_{(2,1)}^{untw.}$ of complex structure moduli $U^j$.

The effective superpotential $\hat W$ arising for non-vanishing 3-form fluxes takes
the following form \TV:
\eqn\TVW{
\hat W={\lambda\over{(2\pi)^2\alpha'}}\ \int_{X_6} G_3\wedge \Omega\ ,}
where $\lambda$ serves to fix the mass dimension to the correct value of 3.
The superpotential gives rise to the standard $F$--term scalar potential
of N=1 supergravity:
\eqn\Vhut{\hat{V}=\hat{K}_{I\ov{J}}\ F^I\ov{F}^{\ov{J}}-3\ e^{\kappa_4^2\hat{K}}\ 
\kappa_4^2\ |\hat{W}|^2\ .}
This is positive semidefinite since the negative contribution in $\hat V$ is
cancelled by the contribution of the K\"ahler moduli $T^i$. 
The integral \TVW\ has been worked out for the orbifold compactifications $X_6$  
we are discussing here in {\it Ref.} \LustFI:
\eqn\WhT{\eqalign{{1\over \lambda}\ \hat{W}&=(a^0-Sc^0)\ U^1U^2U^3-\lf\{(a^1-Sc^1)\ U^2\ U^3+
(a^2-Sc^2)\ U^1\ U^3\ri.\cr
&\lf.+(a^3-Sc^3)\ U^1\ U^2\ri\}-\sum_{i=1}^3(b_i-Sd_i)\ U^i-(b_0-Sd_0)\ .}}
For our purposes, only the supersymmetry breaking $F$--terms $F^S$ and $F^{T^i}$, 
which are proportional to $G_{(3,0)}$ or $G_{(0,3)}$, respectively are relevant \LustFI:
\eqn\ftermsT{\eqalign{
\ov{F}^{\ov{S}}&=(S-\ov{S})^{1/2}\ \prod_{i=1}^3(T^i-\ov{T}^i)^{-1/2}\ 
\prod_{i=1}^3(U^i-\ov{U}^i)^{-1/2}\ \kappa_4^2\ {\lambda\over{(2\pi)^2\alpha'}}
\int\ov{G}_3\wedge \Omega\cr
&=\lambda\ \kappa_4^2\ (S-\ov{S})^{1/2}\ \prod_{i=1}^3(T^i-\ov{T}^i)^{-1/2}\ 
\prod_{i=1}^3(U^i-\ov{U}^i)^{-1/2}\cr
&\times \{(a^0-\ov{S}c^0)\ U^1U^2U^3-
[(a^1-\ov{S}c^1)\ U^2U^3+(a^2-\ov{S}c^2)\ U^1U^3\cr
&+(a^3-\ov{S}c^3)\ U^1U^2]
-\sum_{i=1}^3(b_i-\ov{S}d_i)\ U^i-(b_0-\ov{S}d_0)\}\ ,\cr
\ov{F}^{\ov{T^i}}&=(S-\ov{S})^{-1/2}\ (T^i-\ov{T^i})^{1/2}\ 
(T^j-\ov{T^j})^{-1/2}\ (T^k-\ov{T^k})^{-1/2}\ 
\prod_{j=1}^3(U^j-\ov{U}^j)^{-1/2}\ \kappa_4^2\ \hat{W}\ .}}

\subsec{$D7$--branes with $f$-flux}

Now we will include $D7$--branes together with their open
string sectors. 
To obtain a chiral spectrum, we
must introduce (magnetic) two--form fluxes $F^j dx^j\wedge dy^j$ on the
internal part of the $D7$--brane world volume. 
Together with the internal $NS$ $B$--field
$b^j$, we have the complete $2$--form flux
$\Fc=\sum\limits_{j=1}^3\Fc^j:=\sum\limits_{j=1}^3(b^j+2\pi \ap F^j)\ dx^j\wedge dy^j$. 
The latter gives rise to the total internal antisymmetric background
\eqn\antisbg{
\pmatrix{0&f^j\cr-f^j&0}\ \ \ ,\ \ \ 
f^j=\fc{1}{(2\pi)^2}\ \int_{T^{2,j}} \Fc^j\ ,}
w.r.t. the $j$--th internal plane.
The $2$--form fluxes $\Fc^j$ have to obey the quantization rule:
\eqn\fluxquant{
m^j\fc{1}{(2\pi)^2\ap}\ \int_{T^{2,j}} \Fc^j=n^j\ \ \ ,\ \ \ n\in \IZ\ ,} 
\ie $f^j=\ap \fc{n^j}{m^j}$. This setup is $T$-dual to intersecting $D6$--branes
in type $IIA$ orientifold compactifications.
In a compact model, all tadpoles arising from the Ramond forms $C_4$ and $C_8$ 
must be cancelled by the $D$--branes or/and by the 3-form fluxes. More concretely, the 
cancellation condition for the tadpole arising from the $RR$ $4$--form $C_4$
is 
\eqn\modcfour{
N_{flux}+2
 \sum_a\ N_a n^1_a\ n^2_a\ n^3_a =32\ ,}
where $N_{flux}$ is given by
\eqn\Nflux{
N_{flux}=\fc{1}{(2\pi)^{4}\ \ap^{2}}\ \int_{X_6} H_3\wedge F_3\ .}
For the $ISD$ $(0,3)$-flux, one finds
$$N_{\rm flux}=4\ |A_0|^2\geq 0,$$
and in the case of the $(3,0)$-flux, 
$$N_{\rm flux}=-4|B_0|^2\leq 0.$$ 
Furthermore, the cancellation conditions for the $8$--form tadpoles yield\foot{These equations
are to be understood, that the numbers $m_a,n_a$ come in orbits of the orbifold group 
$\IZ_N$ or $\IZ_N\times\IZ_M$.}:
\eqn\ceight{\eqalign{
2\ \sum_{a}\ N_a\ m_a^1\ m_a^2 \ n^3_a&=-32\ ,\cr
2\ \sum_{a}\ N_a\ m_a^1\ m_a^3\ n^2_a&=-32\ ,\cr
2\ \sum_{a}\ N_a\ m_a^2\ m_a^3\ n^1_a&=-32\ .}}
The requirement that a branes $a$ with internal $2$--form fluxes $f^j_a$ is supersymmetric 
has the form:
\eqn\susy
{\sum_{j=1}^3\arctan\biggl(\fc{f^j_a}{\im({\cal T}^j)}\biggr)=0\ .}
Furthermore, the condition, that branes $a$ with $2$--form fluxes $f^j_a$ and
$b$ with $2$--form fluxes $f^j_b$ are
mutually supersymmetric is 
\eqn\mutususy{
\sum_{j=1}^3\th_{ab}^j=0\ \mod\ 2\ ,}
with the relative ``flux'' $\th_{ab}^j$:
\eqn\relativflux{
\th_{ab}^j=\fc{1}{\pi}\ \lf[\ \arctan\lf(\fc{f_b^j}{\im(\Tc^j)}\ri)-
\arctan\lf(\fc{f_a^j}{\im(\Tc^j)}\ri)\ \ri]\ .}
These conditions will fix some of the K\"ahler moduli ${\cal T}^j$.

Note that in the locally supersymmetric MSSM, being discussed in the next chapter, the
Ramond tadpole conditions are not satisfied, and hence a model-dependent hidden
sector
will always be required to fulfill these conditions.

\subsec{Open string low-energy effective action and soft terms}

The low--energy effective action for the massless open string sector of the
$D3/D7$--branes was computed by calculating string scattering amplitudes among open
string matter fields on the $D$-branes and bulk moduli fields  \doubref\LMRS\LustFI.
Specifically the charged matter fields $C$ enter the
K\"ahler potential at quadratic order as (for large K\"ahler moduli, which corresponds to the 
supergravity approximation under consideration): 
 \eqn\KK{\eqalign{
K(M,\ov M, C, \ov C)=&\hat{K}(M, \ov M)+\sum_{a}\sum_{j=1}^3\sum_{i=1}^3 
G_{C^{7_a,j}_i\ov C^{7_a,j}_i}(M,\ov M)\ C^{7_a,j}_i\ov C^{7_a,j}_i\cr
&+\sum_{a\neq b} G_{C^{7_a7_b}\ov C^{7_a7_b}}(M,\ov M)\ C^{7_a7_b}\ \ov C^{7_a7_b}+\Oc(C^4)\ .}}
Here, $\hat{K}(M, \ov M)$ is the closed string moduli K\"ahler potential \Khut, discussed before.
The open string moduli fields $C$ summarize both untwisted $D7$--brane moduli $C^{7,j}_i$ 
and twisted matter fields $C^{7_a7_b}$. The fields $C^{7,j}_i$ 
account for the transverse $D7$--brane positions $C^{7,j}_j$ on the $j$--th subplane
and for the Wilson line moduli $C^{7,j}_i\ ,\ i\neq j$  on the $D7$--brane world volume. 
On the other hand, the fields $C^{7_a7_b}$ represent twisted matter fields originating 
from strings stretched between two stacks of $D7$--branes $a$ and $b$.
We have only displayed the $D7$--brane sector, as the $D3$--brane sector follows from
the latter by taking the limits $f^j\ra\infty$.
Furthermore, the holomorphic superpotential $W$ takes the form:
\eqn\super{\eqalign{
W(M, C)&=\hat{W}(M)+\sum_{a=1}^3 C^{7_a}_1 C_2^{7_a} C_3^{7_a}+\sum_{a,b,j} d_{abj}\ 
C_j^{7_a} C^{7_{a}7_b} C^{7_a 7_b}\cr
&+C^{7_17_2}C^{7_37_1}C^{7_27_3}+\sum_{I,J,K} Y_{IJK}(U^i)\ C_IC_JC_K+ \Oc(C^4)\ .}}
Again, $\hat{W}(M)$ is the closed string superpotential \TVW, discussed before.
Finally, the coupling of the (closed string) moduli to the gauge fields is described by the
gauge kinetic functions. For the gauge fields living on the $D7$--branes,
wrapped around the $4$--cycle $T^{2,k}\times T^{2,l}$, these functions are given by 
\doubref\LMRS\LustFI
\eqn\gauge{
f_{D7_j}(S,T^j)=|m^k m^l|\lf(T^j-\ap^{-2} f^kf^lS \ri) \ \ \ , \ \ \
(j,k,l)=\overline{(1,2,3)}\ ,}
$m^k, m^l$ being the wrapping numbers.

As we will see in the next chapter, the MSSM--like model entirely consists
of three stacks of $D7$--branes, one being equipped with non-trivial $f$--flux. 
The MSSM matter fields correspond to twisted open string sectors which preserve 
N=1 supersymmetry ($1/4$ BPS sectors).
For those  two stacks of $D7$--branes $a$ and $b$, which wrap different
$4$--cycles, in each plane there is always a non--vanishing relative ``flux'' $\th_{ab}^j$, given
in \eqq \relativflux.

In that case the matter field K\"ahler metric describing a 1/4 BPS sector 
is given by the following expression\foot{
For $\beta,\gamma=0$ this expression agrees with the two results \eqqs (5.22) and (5.25) of \LMRS\ 
after transforming the latter into the Einstein frame. The latter have been extracted  
from a certain three--point
and four--point amplitude in Type $IIA$. However, to completley fix the moduli 
dependence on $T^i$ ($U^i$ in \ta), \ie to fix the constants $\beta,\gamma$,  
one has to calculate a 
four--point disk amplitude involving two twisted matter fields and two K\"ahler moduli $T^i$ (two 
complex structure moduli $U^i$ in \ta) \wip. 
Note, that as in the heterotic case, these constants cannot be determined from factorizing the 
four--twist correlator on the disc. Moreover, results from the heterotic string suggest, that
$\beta,\gamma\neq0$. Recently, in 
{\it Refs.} \recently\ soft--terms have been calculated with assuming $\beta,\gamma=0$.} 
\doubref\LMRS\LustFI:
\eqn\siebensieben{
G_{C^{7_a7_b}\ov C^{7_a7_b}}=\kappa_4^{-2}\ (S-\ov S)^{-\fc{1}{4}+\fc{3\beta}{2}+\gamma}
\prod_{j=1}^3 (T^j-\ov T^j)^{-\fc{1}{4}-\fc{\beta}{2}-\gamma(1-\theta_{ab}^j)}\ 
(U^j-\ov U^j)^{-\th_{ab}^j}\ 
\sqrt\fc{\Gamma(\th_{ab}^j)}{\Gamma(1-\th_{ab}^j)}\ .}

On the other hand, for twisted open string states from the $1/2\ BPS$--sector, the metric takes
a different form:
\eqn\halfBPS{
G_{C^{7_27_3}\ov C^{7_27_3}}={-\kappa_4^{-2}\over(S-\ov S)^{1/2}
(T^1-\ov T^1)^{1/2}}{1\over(U^2-\ov U^2)^{1/2}(U^3-\ov U^3)^{1/2}}\ .}
The metric for the untwisted matter fields living on the same stack of
$D7$--branes is the following:
\eqn\metricsfield{\eqalign{
G_{C^{7,j}_i\ov C^{7,j}_i}&={-\kappa_4^{-2}\over {(U^i-\ov U^i)\ (T^k-\ov{T}^k)}}\ 
{{|1+i\tilde{f}^k|}\over{|1+i\tilde{f}^i|}}\ ,\cr
G_{C^{7,j}_j\ov C^{7,j}_j}&={-\kappa_4^{-2}\over{(U^j-\ov U^j)\ (S-\ov{S})}}\ 
|1-\tilde{f}^i\tilde{f}^k|\ ,\quad i\neq k\neq j\ .}}
Once we have calculated the corresponding Riemann tensors, we are ready to
write down the scalar mass terms. 
For K\"ahler manifolds the components of the Riemann curvature tensor are given as follows:
\eqn\Ri{
R_{M\ov N i\ov i} = K_{C_i\ov C_iM\ov N}-K_{C_iM\ov C_k}\ G^{\ov C^kC^k}\ 
K_{\ov C_i\ov NC_k}\ .}
For the untwisted matter fields of the stacks without $f$-flux, the curvature
tensors take a particularly simple form, but as the expressions are more
cumbersome for stack 1 and for the twisted case, the reader is referred to
appendix \appA\  for details.

The general formula for the scalar mass terms ist the following:
\eqn\msoft{\eqalign{
(m^{7,j}_{i\ov i})^2&=\kappa_4^2\ [\ (\ |m_{3/2}|^2+\kappa_4^2\hat{V}\ )\ G_{C^{7,j}_i\ov
C^{7,j}_i}-\sum_{M,N}F^M\ov F^{\ov N}R^{7,j}_{M\ov N i\ov i}\ ]\ ,\cr
(m^{7a7b})^2&=\kappa_4^2\ [\ (\ |m_{3/2}|^2+\kappa_4^2\hat{V}\ )\ G_{C^{7a7b}\ov
C^{7a7b}}-\sum_{M,N}F^M\ov F^{\ov N}R^{7a7b}_{M\ov N}\ ]\ ,}}
where $M,\ N$ run over $S,\ T^i,\ U^i$. 
The trilinear coupling is
\eqn\Aijk{\eqalign{
A_{IJK}&=i\prod_M(M-\ov M)^{-1}{\kappa_4^2\lambda\over
(2\pi)^2\alpha'}\left[Y_{IJK}\int G_3\wedge \ov\Om+3\ Y_{IJK}\int \ov G_3\wedge \ov
\Om\right.\cr
&+\left.\sum_i\int \ov G_3\wedge \ov\om_{A_i}\ (Y_{IJK}-(U^i-\ov U^i)\ \p_{U^i}Y_{IJK})\right]\cr
&-i\prod_M(M-\ov M)^{-1/2}\ F^N\ G^{\ov C_IC_I}\ \partial_N G_{C_I(\ov C_I}Y_{JK)I}\ .}}
The term $-\sum_i\int \ov G_3\wedge \ov\om_{A_i}(U^i-\ov U^i)\p_{U^i}Y_{IJK}$
appears, because general $Y_{IJK}$ may depend on the complex structure moduli.
For the gaugino masses we need the gauge kinetic functions \gauge.
Through them, we obtain the gaugino masses:
\eqn\gauginoexpl{
m_{g, D7j}=F^S\ {{-\alpha'^{-2}f^kf^l}\over{(T^j-\ov{T}^j)-\alpha'^{-2}f^kf^l(S-\ov{S})}}+
F^{T^j}\ {{1}\over{(T^j-\ov{T}^j)-\alpha'^{-2}f^kf^l(S-\ov{S})}}\, .}

\newsec{Soft terms for the MSSM from a local $D7$--brane construction}

\subsec{Local MSSM construction with three generations}

The locally supersymmetric MSSM-model was orginally formulated \CIM\ in terms
of three stacks of intersecting $D6$--branes in type $IIA$ compactifications.
The non--trivial intersection angles are necessary in order to obatin three
generations of chiral quark and lepton superfields.
As emphasized before, we are discussing supersymmetry breaking in the context
of \tb orientifold compactifiactions. Hence we have to consider
the $T$-dual version of the MSSM $D6$--brane configuration of \CIM.
This $T$-duality transformation is very easy to find, one has to
perform three $T$-duality transformations with respect to either an $x$--
or an $y$--direction in each of the three two--dimensional subtori $T^{2,i}$.
After the $T$-duality transformation, all three stacks become $D7$--branes, which are 
wrapped around the $4$--cycles $T^{2,1}\times T^{2,2}$, $T^{2,1}\times T^{2,3}$ or
$T^{2,2}\times T^{2,3}$, \ie being transversal to $T^{2,3}$, $T^{2,2}$, $T^{2,1}$, respectively.
The last stack is equipped with non-trivial
 $f$--flux, which is required for a realistic model with three generations of chiral matter fields.
Specifically, the locally supersymmetric MSSM is built 
by three stacks of $D7$--branes with the $f$--flux quantum numbers ($f^j=\ap\fc{n^j}{m^j}$) 
displayed in Table 1.
\vskip0.5cm
{\vbox{\ninepoint{
$$
\vbox{\offinterlineskip\tabskip=0pt
\halign{\strut\vrule#
&~$#$~\hfil
&\vrule#
&~$#$~\hfil
&\vrule#&\vrule#
&~$#$~\hfil
&~$#$~\hfil
&~$#$~\hfil
&~$#$~\hfil
&~$#$~\hfil
&\vrule#
&~$#$~\hfil
&\vrule#
&~$#$~\hfil
&\vrule#  
\cr
\noalign{\hrule}
&
{\rm Stack}
&&
{\rm Gauge\ group}
&&&
(m^1,n^1)
&&
(m^2,n^2)
&&
(m^3,n^3)
&&
N_a
&
\cr
\noalign{\hrule}
\noalign{\hrule}
&
1
&&
U(4)
&&&
-
&&
(1,g)
&&
(-1,g)
&&
4
&
\cr
&
2
&&
SU(2)
&&&
(1,0)
&&
-
&&
(-1,0)
&&
2
&
\cr
&
3
&&
SU(2)
&&&
(1,0)
&&
(-1,0)
&&
-
&&
2
&
\cr
\noalign{\hrule}}}$$
\vskip-6pt
\centerline{\noindent{\bf Table 1:}
{\sl MSSM $D7$--brane configuration with $f$--flux numbers $(m^j,n^j)$.}}
\vskip10pt}}}
\vskip-0.5cm \ \br
The corresponding gauge group is
$G=U(4)\times U(2)\times U(2)$.\foot{Note that in some orbifold models,
the $N_a$ will take  values different from
those in Table 1, if the $D$--branes are fixed
under the orbifold group $\IZ_N\times \IZ_N$ and $\Omega I_n$; e.g.
for the $\IZ_2\times \IZ_2$ orientifold, 
$N_1=8$
because the corresponding gauge group is broken to $U(N_1/2)$ by
the orbifold symmetry (see e.g. \MaSh).} 
Only stack 1 carries non-trivial $f$--flux.
For $g=3$, it contains three chiral generations 
of supersymmetric MSSM matter fields, namely
left-handed matter fields in the representations
$3(\underline 4,\underline 2, \underline 1)$ from
open strings stretching between the (12)-branes,
3 right-handed matter fields  in the representations
$3(\underline 4,\underline 1, \underline 2)$ from the (13) open string sector 
and a Higgs multiplet in the representations
 $(\underline 1,\underline 2, \underline 2)$ from the (23)-sector.
By pulling apart the first stack of branes into a stack of 3
$D$--branes plus one $D$--brane, the $SU(4)$
gauge
group is Higgsed to $SU(3)\times U(1)_{B-L}$,
and the matter fields decompose into the known SM
representations of quarks and leptons.

The supersymmetry condition \susy\ applied on stack 1 yields the following
requirement for the K\"ahler moduli:
\eqn\susya{
{\cal T}^2={\cal T}^3\equiv\cal T\ .}
For stacks 2 and 3, it is fulfilled trivially, as those stacks do not carry
$f$-flux.
These three stacks of $D7$--branes will be a subsector in any concrete global model
that satisfies the Ramond tadpole
conditions \modcfour\ and \ceight\ by the addition of fluxes and
some additional hidden sectors 
(see e.g. the model of \MaSh\ which includes also
supersymmetric or non-supersymmetric 3--form fluxes).
However, as already emphasised, these three stacks of $D7$--branes alone do not
satisfy the tadpole conditions. Plugging in the $f$--flux numbers of the MSSM-branes
into equations  \modcfour\ and \ceight,
the so far uncancelled tadpoles must be eliminated by the hidden sector branes
together with the 3-form fluxes. For $N_1=4$, $N_2=N_3=2$ (see also the
previous footnote), 
the hidden
$D$--branes must satisfy the following Ramond tadpole conditions conditions
($b$ runs over all hidden branes): 
\eqn\hidden{\eqalign{
 N_{flux}+2
 \sum_a\ N_b^h\ n^1_b\ n^2_b\ n^3_b &=-40\ ,\cr
2\ \sum_{b}\ N_b^h\ m_b^1\ m_b^2 \ n^3_b&=-28\ ,\cr
2\ \sum_{b}\ N_b^h\ m_b^1\ m_b^3\ n^2_b&=-28\ ,\cr
2\ \sum_{b}\ N_b^h\ m_b^2\ m_b^3\ n^1_b&=-24\ .}}
How these conditions will be eventually met depends on the concrete compact model.
Generically, as it has been recently emphasized in {\it Ref.} \MaSh, a setup of only
$D3$-- and $D7$--branes is not enough. One should add magnetized
$D9$--$\overline{D9}$--branes.

\subsec{Soft terms for the local MSSM construction}

As we have seen in the last section, the only soft terms that are generated by turning on 
3--form flux in our brane setup are the scalar mass terms and the trilinear couplings, plus 
gravitino and gaugino masses. Due to the fact that we break supersymmetry from $N=1$, 
no fermionic mass terms and no $B$-terms appear.

The open strings representing the untwisted matter fields are those which 
have both ends on the same brane stack and originate from  
dimensional reduction of the $D=10$ gauge field. In the case of a $D7$--brane, we 
have two complex Wilson line moduli and one complex scalar 
which describes the transverse position of the brane. These fields do not correspond to any 
MSSM--fields and must, for the model to be realistic, acquire large masses by some additional 
effect.

The open strings which are interesting for us from the Standard Model point of view come 
from the twisted sector, which consists of fields living at the intersections of the three 
brane stacks. The massless twisted $R$--sector gives rise to chiral fermions in the 
bifundamental representation, while the massless scalar matter fields stem from the 
twisted $NS$--sector. 
The matter fields living at the $(12)$-intersection form the left-handed part of the spectrum, 
the fields living at the $(13)$-intersection form the right-handed part of the spectrum, while 
the fields living at the $(23)$-intersection form the Higgs multiplet. 
Expressed in $N=1$ language, the above matter fields form chiral multiplets, consisting of a 
complex scalar, a spinor and an auxiliary scalar. It is the scalar component, that acquires 
mass through SUSY-breaking.
So what we calculate in the twisted sector are squark and slepton masses, as well as the 
mass that appears in the Higgs potential.
The contribution to the soft terms that is sensitive to the $f$-fluxes comes from the
open string matter metrics, which in turn receive their $f$-form flux dependence from the 
mixed boundary conditions. In our specific setup, we have to deal with the
metric for the untwisted $D7$-brane matter fields and the metric for the twisted
$D7$--brane matter fields, where the fields come from two different stacks of
branes. We use the metrics that were computed in \LustFI\  and plug in the
specific values of $f^j=\alpha'{n^j\over m^j}$ 
of our model as given in Table 1.

We will first examine the untwisted matter metrics. Only stack 1 carries
non-trivial $f$-flux
\eqn\stackoneu{\eqalign{
G^{7,1}_{C_1\ov C_1}&={-\kappa_4^{-2}\over(U^1-\ov U^1)(S-\ov S)}|1+a_2a_3|\ ,\cr
G^{7,1}_{C_2\ov C_2}&={-\kappa_4^{-2}\over(U^2-\ov U^2)(T^3-\ov T^3)}{|1-i
a_3|\over|1+ia_2|}\ ,\cr
G^{7,1}_{C_3\ov C_3}&={-\kappa_4^{-2}\over(U^3-\ov U^3)(T^2-\ov T^2)}{|1+i
a_2|\over|1-ia_3|}\ ,}}
where we define $a_2={\alpha'g\over {\rm Im}{\cal T}^2},\ a_3={\alpha'g\over
{\rm Im}{\cal T}^3}$.
The other two brane stacks have vanishing $f$-flux and the metrics reduce to the simple
form (e.g. stack 2):
\eqn\stacktwou{\eqalign{
G^{7,2}_{C_1\ov C_1}&={-\kappa_4^{-2}\over(U^1-\ov U^1)(T^3-\ov T^3)}\cr
G^{7,2}_{C_2\ov C_2}&={-\kappa_4^{-2}\over(U^2-\ov U^2)(S-\ov S)}\cr
G^{7,2}_{C_3\ov C_3}&={-\kappa_4^{-2}\over(U^3-\ov U^3)(T^1-\ov T^1)}.
}}
For stack 3, we have the same form with indices 2 and 3 interchanged.
Now we turn to the twisted matter metrics. The matter fields between stacks 1
and 2 and between stack 1 and 3 form a $1/4\ BPS$ sector. Their metrics are:
\eqn\twisteda{\eqalign{
G_{C^{7_17_2}\ov C^{7_17_2}}&=2i\ \kappa_4^{-2}\ (S-\ov S)^{-\fc{1}{4}+\fc{3\beta}{2}+\gamma}
\ \fc{\Gamma[\h-\fc{1}{\pi}\arctan(a_2)]}{\pi^{1/2}(1+a_2^2)^{1/4}}\ 
\fc{\Gamma[\fc{1}{\pi}\arctan(a_3)]}{({\pi\over a_3}\sqrt{1+a_3^2})^{1/2}}\cr 
&\hskip-0.85cm\times (T^1-\ov T^1)^{-\fc{1}{4}-\fc{\beta}{2}-\fc{3}{2}\gamma}\ 
(T^2-\ov T^2)^{-\fc{1}{4}-\fc{\beta}{2}-\fc{\gamma}{2}-\fc{\gamma}{\pi}\arctan(a_2)}\ 
(T^3-\ov T^3)^{-\fc{1}{4}-\fc{\beta}{2}-\gamma+\fc{\gamma}{\pi}\arctan(a_3)}\cr
&\hskip-0.85cm\times(U^1-\ov U^1)^{1/2}\ (U^2-\ov U^2)^{-[\h-\fc{1}{\pi}\arctan(a_2)]}\ 
(U^3-\ov U^3)^{-\fc{1}{\pi}\arctan(a_3)}\ ,\cr
G_{C^{7_17_3}\ov C^{7_17_3}}&=-2\ \kappa_4^{-2}\ (S-\ov S)^{-\fc{1}{4}+\fc{3\beta}{2}+\gamma}\ 
\fc{\Gamma[\h+\fc{1}{\pi}\arctan(a_3)]}{\pi^{1/2}(1+a_3^2)^{1/4}}\ 
\fc{\Gamma[-\fc{1}{\pi}\arctan(a_2)]}{({\pi\over a_2}\sqrt{1+a_2^2})^{1/2}}\cr
&\hskip-0.85cm\times (T^1-\ov T^1)^{-\fc{1}{4}-\fc{\beta}{2}-\fc{3}{2}\gamma}\ 
(T^2-\ov T^2)^{-\fc{1}{4}-\fc{\beta}{2}-\gamma-\fc{\gamma}{\pi}\arctan(a_2)}\ 
(T^3-\ov T^3)^{-\fc{1}{4}-\fc{\beta}{2}-\fc{\gamma}{2}+\fc{\gamma}{\pi}\arctan(a_3)}\cr
&\hskip-0.85cm\times (U^1-\ov U^1)^{1/2}(U^2-\ov
U^2)^{\fc{1}{\pi}\arctan(a_2)}(U^3-\ov U^3)^{-[\h+\fc{1}{\pi}\arctan(a_3)]}\ .}}
The matter fields between stacks 2 and 3, which do not carry fluxes are $1/2\
BPS$ states and the metric has a different form, as given in \halfBPS.

For simplicity, and as these fluxes are generic for all orbifold groups, only
$(0,3)$- and $(3,0)$-fluxes are turned on.

\vskip0.5cm
\noindent
{\it(i) Scalar mass terms}

First, we examine the untwisted case. With the curvature tensors given in appendix~\appA\  
and
\eqn\Y{
Y=(S-\ov S)\ \prod_{j=1}^3 (T^j-\ov T^j)\ (U^j-\ov U^j)\ ,}
we come to the following result:

\noindent
Stack 1:
$$\eqalign{
(m^{7,1}_{1\ov 1})^2&={\lambda^2\kappa_4^6\over (2\pi)^4\alpha'^2}\ \fc{G_{C_1\ov C_1}^{7,1}}{|Y|}
\cr
&\times\lf\{\ \lf(\ 1-\fc{\tii\ \tiii\ (2\ g^2\ap^2+\tii\ \tiii)}{(g^2\ap^2+\tii\ \tiii)^2}\ 
\ri)\ \lf|\ \int\ov G_3\wedge \Omega\ \ri|^2\ri.\cr
&+\fc{g^2\ap^2\ \tii\tiii}{(g^2\ap^2+\tii\tiii)^2}
\ \lf(\int\ov G_3\wedge \Omega\ \int \ov G_3\wedge \ov\Omega+ c.c.\ri)  \cr  
&+\lf. \lf(\ 1-\fc{g^2\ap^2\ (g^2\ap^2+2\tii\ \tiii)}{(g^2\ap^2+\tii\ \tiii)^2}\ \ri)\ 
\lf|\ \int G_3\wedge \Omega\ \ri|^2\ \ri\}\ ,}$$
\eqn\msoftstacka{\eqalign{
(m^{7,1}_{2\ov 2})^2&={\lambda^2\kappa_4^6\over (2\pi)^4\alpha'^2}\ \fc{G_{C_2\ov C_2}^{7,1}}{|Y|}
\cr
&\times\lf\{\lf(\ 1+\fc{g^4\ap^4\ [(\tii)^2-(\tiii)^2]\ [2g^2\ap^2+(\tii)^2+
(\tiii)^2]}{2\ [g^2\ap^2+(\tii)^2]^2\ [g^2\ap^2+(\tiii)^2]^2}\ \ri)\ 
\lf|\ \int\ov G_3\wedge \Omega\ \ri|^2\ri.\cr
&-\h\ g^2\ap^2\lf(\fc{(\tii)^2}{[g^2\ap^2+(\tii)^2]^2}-\fc{(\tiii)^2}{[g^2\ap^2+(\tiii)^2]^2}
\ri)\ \lf(\int\ov G_3\wedge \Omega\ \int \ov G_3\wedge \ov\Omega+ c.c.\ri)  \cr  
&+\lf. \lf(\h-\fc{(\tii)^4}{2\ [g^2\ap^2+(\tii)^2]^2}-
\fc{g^4\ap^4+2g^2\ap^2\ (\tiii)^2}{2\ [g^2\ap^2+(\tiii)^2]^2}\ri)\ 
\lf|\ \int G_3\wedge \Omega\ \ri|^2\ \ri\}\ .}}
The mass 
$(m^{7,1}_{3\ov 3})^2$ is obtained by using $G^{7,1}_{C_3\ov C_3}$ instead of 
$G^{7,1}_{C_2\ov C_2}$ and interchanging $\im\Tc^2$ and $\im\Tc^3$, otherwise is has the same 
structure as $(m^{7,1}_{2\ov 2})^2$.

\noindent
Stack 2:
\eqn\msoftstackb{\eqalign{
(m^{7,2}_{1\ov 1})^2&={\lambda^2\ \kappa_4^6\over (2\pi)^4\alpha'^2}\ 
\fc{G_{C_1\ov C_1}^{7,2}}{|Y|}\ 
\lf|\ \int\ov G_3\wedge \Omega\ \ri|^2\ ,\cr
(m^{7,2}_{2\ov 2})^2&={\lambda^2\ \kappa_4^6\over (2\pi)^4\alpha'^2}\ 
\fc{G_{C_2\ov C_2}^{7,2}}{|Y|}\ 
\lf|\ \int G_3\wedge \Omega\ \ri|^2\ ,\cr
(m^{7,2}_{3\ov 3})^2&={\lambda^2\ \kappa_4^6\over (2\pi)^4\alpha'^2}\ 
\fc{G_{C_3\ov C_3}^{7,2}}{|Y|}\ 
\lf|\ \int\ov G_3\wedge \Omega\ \ri|^2\ .}}
The mass terms for stack 3 are obtained by interchanging the indices 2 and 3.

In stack 1, which is the stack carrying $f$-fluxes, all the flux components
appear, they even mix. For the stacks without $f$-flux, the general formula
\msoft\ simplifies drastically.
Remarkably, in stacks 2 and 3 the mass term concerning the two-cycle which is
not wrapped by the respective stack of 7-branes differs from the others in its
dependence on the 3-form flux: While the other mass terms contain the
$(3,0)$-flux piece, $m^{7,3}_{3\ov 3}$ and $m^{7,2}_{2\ov 2}$  contain the
$(0,3)$-flux piece. 

Now, we look at the twisted case. 
The values of $\theta_{ab}^j$ (which is
the relative angle between the brane stacks in the $T$-dual picture) are the
following for our specific model:
\eqn\thetas{\eqalign{
\theta^1_{12}&=-\h,\quad \theta^2_{12}=\h-{1\over \pi}\arctan(a_2),\quad
\theta^3_{12}={1\over \pi}\arctan(a_3)\ ,\cr
\theta^1_{13}&=-\h,\quad \theta^2_{13}=-{1\over \pi}\arctan(a_2),\quad
\theta^3_{13}=\h+{1\over \pi}\arctan(a_3)\ ,\cr
\theta^1_{23}&=0,\quad \theta^2_{23}=-\h,\quad \theta^3_{23}=\h\ .}}
Again, we get a simple form for the
$(23)$-sector, whereas for the $(12)$- and $(13)$-sectors with nontrivial
$\theta_{ab}$, the case is more complicated (see appendix \appA\ for details):
\eqn\mtwistedab{\eqalign{
(m^{7a7b})^2&={\lambda^2\kappa_4^6\over (2\pi)^4\alpha'^2}\ \fc{G_{C^{7a7b}\ov C^{7a7b}}}{|Y|}\cr
&\times\left[ \lf(\ \fc{3}{4}+\fc{3}{2}\beta+\gamma\ \ri)\ \lf|\int\ov G_3\wedge\Omega\ri|^2
+\lf(\ \fc{1}{4}-\fc{3}{2}\beta-3\gamma+\gamma\sum_j\theta^j_{ab}\ \ri)
\ \lf|\int G_3\wedge\Omega\ri|^2\ri.\cr
&\hskip0.5cm+\fc{\gamma}{2\pi}\ \sum_j s_j\  
\fc{\ap g\ \im\Tc^j}{(\ap g)^2+(\im\Tc^j)^2}\ \ 
\lf(-2\lf|\int G_3\wedge\Omega\ri|^2+(\int\ov G_3\wedge\Omega \int \ov G_3\wedge\ov\Omega
+c.c.)\ri)\cr
&-\fc{1}{4}\sum_{i=2,3}\fc{g\ap}{\pi\ [(g\alpha')^2+(\im\Tc^i)^2]^2}\cr
&\times \lf\{s_i\ \lf(\gamma\ \ln(T^i-\ov T^i)-\ln(U^i-\ov U^i)+
\h\ [\ \psi_0(\theta^i_{ab})+\psi_0(1-\theta^i_{ab})]\ \ri)\ \ri.\cr
&\hskip0.5cm\times 
\lf[\ (\im\Tc^i)^3\ \lf(\lf|\int\ov G_3\wedge\Omega\ri|^2-3\lf|\int G_3\wedge\Omega\ri|^2
+(\int\ov G_3\wedge\Omega \int \ov G_3\wedge\ov\Omega+c.c.)\ri)\ri.\cr
&\lf.\hskip1cm+
(\ap g)^2\ \im\Tc^i\ \lf(3\lf|\int\ov G_3\wedge\Omega\ri|^2-\lf|\int G_3\wedge\Omega\ri|^2-
(\int\ov G_3\wedge\Omega \int \ov G_3\wedge\ov\Omega+c.c.)\ri)\ri]\cr
&+\h\ {\ap g\over \pi}\ (\im\Tc^i)^2\ [\ \psi_1(\theta^i_{ab})-\psi_1(1-\theta^i_{ab})\ ]\cr
&\hskip0.5cm\times\lf.\lf.
\lf(\lf|\int\ov G_3\wedge\Omega\ri|^2+\lf|\int G_3\wedge\Omega\ri|^2
-(\int\ov G_3\wedge\Omega \int \ov G_3\wedge\ov\Omega+c.c.)\ri)\  \ri\}\ri]\ .}}
with $(a,b)=(1,2)$ or $(1,3)$. 
In this case, the 2-form flux dependence is very complicated, as the appearance of the 
Gamma--function in the original metric already suggested. The 3-form flux appears again with 
a $(3,0)$-part, a $(0,3)$-part and a combination of the two.

For $(a,b)=(2,3)$, the $1/2\  BPS$--case that corresponds to the Higgs multiplet, we get
\eqn\mtwisted{
(m^{7_27_3})^2={\lambda^2\kappa_4^6\over (2\pi)^4\alpha'^2}\ 
{G_{C^{7_27_3}\ov C^{7_27_3}}\over2 |Y|}\ 
\left(\ \lf|\int G_3\wedge \Omega\ \ri|^2+\lf|\int\ov G_3\wedge
\Omega\ \ri|^2\ \ri)\ ,}
which is a lot simpler as the world--volume $2$--form $f$--flux does not enter.

\vskip0.5cm
\noindent
{\it(ii) Trilinear couplings}

Now, we will examine the trilinear coupling, at least in the example of the untwisted matter 
fields. In the case of all three fields
living on the same stack of branes, the general formula \Aijk\ simplifies
considerably, as $Y_{ijk}=\epsilon_{ijk}$:
\eqn\trilineara{\eqalign{
A_{ijk}^{7,1}&=\epsilon_{ijk}\prod|M-\ov M|^{-1}{\kappa_4^2\lambda\over
(2\pi)^2\alpha'}\left\{\left[{-(\alpha'g)^2\over (\alpha'g)^2+\im\Tc^2\im\Tc^3}-
{(\alpha'g)^2\over (\alpha'g)^2+(\im\Tc^2)^2}\right.\right.\cr
&\left.-{(\alpha'g)^2\over (\alpha'g)^2+(\im\Tc^3)^2}\right] \int G_3\wedge \ov\Omega\, 
+\left[1+{(\alpha'g)^2\over (\alpha'g)^2+\im\Tc^2\im\Tc^3}\right.\cr
&\left.\left.+{(\alpha'g)^2\over (\alpha'g)^2+(\im\Tc^2)^2}+{(\alpha'g)^2\over (\alpha'g)^2+
(\im\Tc^3)^2}\right]\int \ov G_3\wedge \ov\Omega\ \right\},\cr
A_{ijk}^{7,2}&=A_{ijk}^{7,3}=\epsilon_{ijk}\prod|M-\ov M|^{-1}{\kappa_4^2\lambda\over
(2\pi)^2\alpha'}\int\ov G_3\wedge\ov\Omega\ .}}
For the stacks with vanishing 2-form flux, only the $(0,3)$-part contributes,
as opposed to the result obtained for untwisted matter fields on a
$D3$--brane, where only the $(3,0)$--part contributes. The coupling for stack
1 remains complicated due to the non-trivial moduli dependence of the metric.

\vskip0.5cm
\noindent
{\it(iii) Gaugino masses}

Last, but not least, we also give the gaugino masses for our specific model.
The gaugino masses are derived from the gauge kinetic function given in
\gauge\  via $m_{g,j}=F^M\partial_M\log({\rm Im}f_{D7,j})$.
For our model, we have:
\eqn\gaugekin{\eqalign{
f_{1}&=T^1+g^2S\ ,\cr
f_{j}&=T^j\ \ \ ,\ \ \ j=2,3\ .}}
This gives us
\eqn\gauginos{\eqalign{
m_{g,1}&={F^{T^1}+g^2F^S\over (T^1-\ov T^1)+g^2(S-\ov S)}\ ,\cr
m_{g,j}&={F^{T^j}\over(T^j-\ov T^j)},\ j=2,3\ .}}

\subsec{Concrete Example}

We will get even more specific now. We will look at the soft terms for our
MSSM-model with the SUSY--condition \susya\ enforced, which leads to $T^2=T^3\equiv
\tilde T\,$. Thanks to the supersymmetry
condition, we are able to eliminate the string basis moduli completely
in the following. Furthermore, we turn on a specific 3-form flux consisting of a $(3,0)$-part and
a $(0,3)$-part obtained for $U^1=U^2=U^3=S=i$ and take $g=3$, which results in three fermion 
generations.

A flux solution with $(0,3)$- and $(3,0)$-component for  $U^1=U^2=U^3=S=i$ is
\eqn\fluxsol{\eqalign{
{1\over (2\pi)^2\alpha'}\ G_{(0,3)+(3,0)}&=(b_3 -i d_3) \alpha_0 + (-b_0 +i d_0) \alpha_1 + 
(-b_0 +id_0) \alpha_2 + (-b_0 +i d_0) \alpha_3\cr
& + (b_0 - i d_0) \beta_0 + (b_3 - i d_3)
\beta_1 + (b_3 - i d_3) \beta_2 + (b_3 - i d_3) \beta_3\ .}}
For the real coefficients $b_0,\ b_3,\ d_0,\ d_3$, any integer number can be
chosen. To avoid possible complications with flux quantization, we take the
coefficients to be multiples of 8, though.
Expressed in the complex basis, this flux reads
\eqn\fluxsolc{
{1\over (2\pi)^2\alpha'}\ G_{(0,3)+(3,0)}=\h\ (b_3+d_0+i(b_0-d_3))\ \om_{A_0}+
\h\ (b_3-d_0-i(b_0+d_3))\ \om_{B_0}\ .}
We will need the following flux integrals:
\eqn\fluxint{\eqalign{
{1\over (2\pi)^4\alpha'^2}\ \lf|\int G_3\wedge\Omega\ \ri|^2&=16\ [(b_3+d_0)^2+(b_0-d_3)^2]\ ,\cr
{1\over (2\pi)^4\alpha'^2}\ \lf|\int \ov G_3\wedge\Omega\ \ri|^2&=
16\ [(b_3-d_0)^2+(b_0+d_3)^2]\ ,\cr
{1\over (2\pi)^4\alpha'^2}\ \int G_3\wedge\Omega\ \times\int G_3\wedge\ov\Omega\
&=-16\ [b_0^2+b_3^2-d_0^2-d_3^2-2i(b_0d_0+b_3d_3)]\ ,\cr
{1\over (2\pi)^2\alpha'}\ \int \ov G_3\wedge \ov\Omega&=4\ [b_0-d_3+i(b_3+d_0)]\ ,\cr
{1\over (2\pi)^2\alpha'}\ \int G_3\wedge \ov\Omega&=-4\ [b_0+d_3+i(b_3-d_0)]\ .}}

We will again first examine the metrics. The untwisted matter metric for
stack 1 simplifies considerably as the SUSY-condition leads to $a_2=a_3\equiv a$:

\eqn\stackoneus{\eqalign{
G^{7,1}_{C_1\ov C_1}&={i\over 2}\kappa_4^{-2}\ \lf(\fc{1}{S-\ov S}+\fc{9}{T^1-\ov T^1}\ri)\ ,\cr
G^{7,1}_{C_2\ov C_2}&=G^{7,1}_{C_3\ov C_3}={i\over 2}\ 
{\kappa_4^{-2}\over(\tilde T-\ov{\tilde T})}\ .}}
Stack 2 has the following metric:
\eqn\stacktwou{\eqalign{
G^{7,2}_{C_1\ov C_1}&={i\over
2}\ {\kappa_4^{-2}\over(\tilde T-\ov{\tilde T})}\ ,\cr
G^{7,2}_{C_2\ov C_2}&={i\over 2}\ \fc{\kappa_4^{-2}}{S-\ov S}\ ,\cr
G^{7,2}_{C_3\ov C_3}&={i\over
2}\ {\kappa_4^{-2}\over(T^1-\ov T^1)}\ .}}
The metric for stack 3 can be obtained as usual by interchanging the indices
2 and 3.
The twisted matter metrics simplify as well. With $a=3\sqrt{\fc{S-\ov S}{T^1-\ov T^1}}$, we
obtain:
\eqn\twistedas{\eqalign{
G_{C^{7_17_2}\ov C^{7_17_2}}&=2i\ \kappa_4^{-2}\  
(S-\ov S)^{-\fc{1}{4}+\fc{3}{2}\beta+\gamma}\ 
(T^1-\ov T^1)^{-\fc{1}{4}-\fc{\beta}{2}-\fc{3}{2}\gamma}
(\tilde T-\ov{\tilde T})^{-\h-\beta-\fc{3}{2}\gamma}\cr
&\times \fc{1}{\pi}\ \sqrt{\fc{a}{1+a^2}}\ 
\Gamma\lf[\h-\fc{1}{\pi}\arctan(a)\ri]\ \Gamma\lf[\fc{1}{\pi}\arctan(a)\ri]\ ,\cr
G_{C^{7_27_3}\ov C^{7_27_3}}&=-2\ \kappa_4^{-2}\  (S-\ov S)^{-\fc{1}{4}+\fc{3}{2}\beta+\gamma}\ 
(T^1-\ov T^1)^{-\fc{1}{4}-\fc{\beta}{2}-\fc{3}{2}\gamma}
(\tilde T-\ov{\tilde T})^{-\h-\beta-\fc{3}{2}\gamma}\cr
&\times
\fc{1}{\pi}\ \sqrt{\fc{a}{1+a^2}}\ 
\Gamma\lf[\h+\fc{1}{\pi}\arctan(a)\ri]\ \Gamma\lf[-\fc{1}{\pi}\arctan(a)\ri]\ ,\cr
G_{C^{7_27_3}\ov C^{7_27_3}}&=-\kappa_4^{-2}\,(2i)^{-3/2}{1\over(T^1-\ov T^1)^{1/2}}\ .}}

\vskip0.5cm
\noindent
{\it (i) Scalar mass terms}
\vskip0.2cm

For the untwisted matter fields, the only non-trivial case for the Riemann
tensor is $R^{7,1}_{S\ov S1\ov 1}$, see appendix \appA. The scalar
masses simplify as follows:

Stack 1:
\eqn\sma{\eqalign{
(m_{1\ov 1}^{7,1})^2&=2\ \lambda^2\kappa_4^6\ 
{G_{C_1\ov C_1}^{7,1}\over(S-\ov S)(\tilde T-\ov{\tilde T})^2(T^1-\ov T^1)}\cr
&\times\lf\{\ \lf(\ 1+\fc{(T^1-\ov T^1)\ [18\ (S-\ov S)+T^1-\ov T^1]}
{(S-\ov S)^2\ [ 9\ (S-\ov S)+T^1-\ov T^1]^2}\ \ri)\ 
[(b_3-d_0)^2+(b_0+d_3)^2]\ri.\cr
&+\fc{18}{[9\ (S-\ov S)+T^1-\ov T^1]^2}\ (b_0^2+b_3^2-d_0^2-d_3^3)\cr
&+\lf.\lf(\ 1+\fc{9\ (S-\ov S)\ [9\ (S-\ov S)+2\ (T^1-\ov T^1)]}
{(T^1-\ov T^1)^2\ [ 9\ (S-\ov S)+T^1-\ov T^1]^2}\ \ri)\ 
[(b_3+d_0)^2+(b_0-d_3)^2]\ \ri\}\ ,\cr
(m_{2\ov 2}^{7,1})^2&=2\ \lambda^2\kappa_4^6\ 
{G_{C_2\ov C_2}^{7,1}\over(S-\ov S)(\tilde T-\ov{\tilde T})^2(T^1-\ov T^1)}\ 
[(b_3-d_0)^2+(b_0+d_3)^2]\ ,\cr
(m_{3\ov 3}^{7,1})^2&=2\ \lambda^2\kappa_4^6\ 
{G_{C_3\ov C_3}^{7,1}\over (S-\ov S)(\tilde T-\ov{\tilde T})^2(T^1-\ov T^1)}\ 
[(b_3-d_0)^2+(b_0+d_3)^2]\ .}}
As apparent already in equation \msoftstacka, as well as from the metrics \stackoneus, 
the 2-form flux dependence drops completely out of $(m^{7,1}_{2\ov 2})^2$ 
and  $(m^{7,1}_{3\ov 3})^2$ for $\im\Tc^2=\im\Tc^3$.

Stack 2:
\eqn\smb{\eqalign{
(m_{1\ov 1}^{7,2})^2&=2\ \lambda^2\kappa_4^6\ {G_{C_1\ov C_1}^{7,2}
\over (S-\ov S)(\tilde T-\ov{\tilde T})^2(T^1-\ov T^1)}\ [(b_3-d_0)^2+(b_0+d_3)^2]\ ,\cr
(m_{2\ov 2}^{7,2})^2&=2\ \lambda^2\kappa_4^6\ {G_{C_2\ov C_2}^{7,2}
\over (S-\ov S)(\tilde T-\tilde{\ov T})^2(T^1-\ov T^1)}\ [(b_3+d_0)^2+(b_0-d_3)^2]\ ,\cr
(m_{3\ov 3}^{7,2})^2&=2\ \lambda^2\kappa_4^6\ {G_{C_3\ov C_3}^{7,2}
\over(S-\ov S)(\tilde T-\ov{\tilde T})^2(T^1-\ov T^1)}\ [(b_3-d_0)^2+(b_0+d_3)^2]\ .}}
The result for stack 3 is obtained by changing the indices.

{From} the above, we immediately see that we are left with a number of
unfixed parameters: the imaginary parts of the two
K\"ahler moduli $T^1$ and $\tilde T$ which are left unfixed, and the four real
parameters describing the 3-form flux. The 2-form flux is fixed by the
requirement that we want to obtain 3 particle generations as in the Standard Model.

In the twisted case, we get the following mass terms for the $1/4\ BPS$-states ($b=2,3$)
\eqn\quarterBPSm{\eqalign{
(m^{7_17_b})^2&=\lambda^2\kappa_4^6\ 
\fc{G_{C^{7_17_b}\ov C^{7_17_b}}}{(S-\ov S)(\tilde T-\ov{\tilde T})^2(T^1-\ov T^1)}\ 
\fc{1}{[9\ (S-\ov S)+T^1-\ov T^1]^2}\ \cr
&\times \lf\{\ 
\lf(m_{SS}-[27\ (S-\ov S)+(T^1-\ov T^1)]\ \Psi^{(0)}_b(a)+\Psi^{(1)}_b(a)\ri)\ 
[(b_3-d_0)^2+(b_0+d_3)^2]\ri.\cr 
&
+\lf(m_{TT}+[9\ (S-\ov S)+3\ (T^1-\ov T^1)]\ \Psi^{(0)}_b(a)+\Psi^{(1)}_b(a)\ri)\ 
[(b_3+d_0)^2+(b_0-d_3)^2]\cr
&\lf.-
\lf(m_{ST}+[9\ (S-\ov S)-\ (T^1-\ov T^1)]\ \Psi^{(0)}_b(a)-\Psi^{(1)}_b(a)\ri)\ 
(b_0^2+b_3^2-d_0^2-d_3^3)\ \ri\}\ ,}}
with:
\eqn\withdef{\eqalign{
\Psi^{(0)}_2(a)&=\fc{3}{2\pi}\ (S-\ov S)^{1/2}(T^1-\ov T^1)^{1/2}\ 
\lf[\ \psi_0\lf(\fc{a}{\pi}\ri)-\psi_0\lf(\h-\fc{a}{\pi}\ri)\ \ri]\ ,\cr
\Psi^{(1)}_2(a)&=\fc{9}{2\pi^2}\ (S-\ov S)\ (T^1-\ov T^1)\ 
\lf[\psi_1\lf(\fc{a}{\pi}\ri)+\psi_1\lf(\h-\fc{a}{\pi}\ri)\ \ri]\ ,\cr
\Psi^{(0)}_3(a)&=-\fc{3}{2\pi}\ (S-\ov S)^{1/2}(T^1-\ov T^1)^{1/2}\ 
\lf[\ \psi_0\lf(-\fc{a}{\pi}\ri)-\psi_0\lf(\h+\fc{a}{\pi}\ri)\ \ri]\ ,\cr
\Psi^{(1)}_3(a)&=\fc{9}{2\pi^2}\ (S-\ov S)\ (T^1-\ov T^1)\ 
\lf[\psi_1\lf(-\fc{a}{\pi}\ri)+\psi_1\lf(\h+\fc{a}{\pi}\ri)\ \ri]\ ,\cr
m_{SS}&=81\ (3-3\beta-2\gamma)\ (S-\ov S)^2+(2-3\beta-2\gamma)\ (T^1-\ov T^1)\ 
[18\ (S-\ov S)+(T^1-\ov T^1)]\ ,\cr
m_{TT}&=(4+3\beta+6\gamma)\ (T^1-\ov T^1)^2+27\ (1+\beta+2\gamma)\ 
(S-\ov S)\ [9\ (S-\ov S)+2\ (T^1-\ov T^1)]\ ,\cr
m_{ST}&=9\ (S-\ov S)\ (T^1-\ov T^1)\ .}}
The $1/2\ BPS$ mass is the following:
\eqn\halfbc{
(m^{7_27_3})^2=2\ \lambda^2\kappa_4^6\ \fc{G_{C^{7_27_3}}}{
(S-\ov S)\ (\tilde T-\ov{\tilde T})^2\ (T^1-\ov T^1)}\ 
\lf(\ b_0^2+b_3^2+d_0^2+d_3^2\ \ri)\ .}

\vskip0.5cm
\noindent
{\it (ii) Trilinear couplings}
\vskip0.2cm
\eqn\trilineara{\eqalign{
A_{ijk}^{7,1}&=\epsilon_{ijk}{1\over|T^1-\ov T^1|}{1\over |\tilde T-\ov{\tilde T}|^2}{\kappa_4^2\lambda\over (2\pi)^2\alpha'}\left\{{3(\alpha'g)^2\over (\alpha'g)^2+(e^{\phi_4}\pi)^4(T^1-\ov T^1)^2(\tilde T-\ov{\tilde T})^2}[b_0+d_3 \right.\cr
&+i(b_3-d_0)]\left.+\left[1+{3(\alpha'g)^2\over (\alpha'g)^2+(e^{\phi_4}\pi)^4(T^1-\ov T^1)^2(\tilde T-\ov{\tilde T})^2}\right][b_0-d_3+i(b_3+d_0)] \right\},\cr
A_{ijk}^{7,2}&=A_{ijk}^{7,3}=\epsilon_{ijk}{1\over 4}{1\over|T^1-\ov T^1|}{1\over |\tilde T-\ov{\tilde T}|^2}{\kappa_4^2\lambda\over (2\pi)^2\alpha'}[b_0-d_3+i(b_3+d_0)].
}}

\vskip0.5cm
\noindent
{\it (iii) Gaugino masses}
\vskip0.2cm

The gaugino masses \gauginos\ have the following form:
\eqn\gauginoexp{\eqalign{
m_{g,1}&=\h\ {\kappa_4^2\lambda\over(S-\ov S)^{1/2}(T^1-\ov T^1)^{1/2}(\tilde T-\ov{\tilde  T})}\cr
&\times {(T^1-\ov T^1)[b_0-d_3+i(b_3+d_0)]-9(S-\ov S)[b_0+d_3+i(b_3-d_0)]
\over(T^1-\ov T^1)+9(S-\ov S)}\ ,\cr
m_{g,2}&=m_{g,3}=\h\ {\kappa_4^2\lambda\over(S-\ov S)^{1/2}(T^1-\ov T^1)^{1/2} 
(\tilde T-\ov{\tilde  T})}\ [\ b_0-d_3+i(b_3+d_0)\ ]\ .}}

\newsec{Goldstino angle and structure of soft--terms}

In this section, we want to rewrite our results from the last section in terms 
of the so--called goldstino angle, for which an isotropic compactification is assumed,
\ie $\Tc^1=\Tc^2=\Tc^3\equiv \Tc$, or $T^1=T^2=T^3\equiv T$, respectively
and $U^j=i$.
Note, that this requirement automatically fulfills the supersymmetry condition \susya.
For an isotropic compactification, the K\"ahler potential \Khut\ boils down to
\eqn\boilK{
\kappa_4^{2}\ \hat K=-\ln(S-\ov S)-3\ \ln(T-\ov T)-3\ln(2i)\ .}
With this we get the two $F$--terms $F^S,\ F^T$, where $F^T$ refers to the overall 
K\"ahler modulus $T$ (\cf section 3):
\eqn\Fterms{\eqalign{
\ov{F}^{\ov{S}}&=(2i)^{-3/2}\ 
(S-\ov{S})^{1/2}\ (T-\ov{T})^{-3/2}\ \kappa_4^2\ {\lambda\over{(2\pi)^2\alpha'}}
\int\ov{G}_3\wedge \Omega\ ,\cr
\ov{F}^{\ov{T}}&=(2i)^{-3/2}\ (S-\ov{S})^{-1/2}\ (T-\ov{T})^{-1/2}\ 
\ \kappa_4^2\ {\lambda\over{(2\pi)^2\alpha'}}
\int {G}_3\wedge \Omega\ .}} 
The goldstino angle $\theta_g$ describes where the source of supersymmetry 
breaking originates: If $\th_g=\fc{\pi}{2},\fc{3\pi}{2},\ldots$, the breaking 
is due to an $F_S$--term from $(3,0)$--form fluxes only
(dilaton--dominated), whereas in the case $\th_g=0,\pi,\ldots$, the breaking is entirely 
due to an $F_T$--term from $(0,3)$--form fluxes. Hence, the ratio
between $F^S$ and $F^T$ can be used to define the goldstino angle
\eqn\tangol{
\tan\theta_g=e^{i(\alpha_T-\alpha_S)}\ {\hat K_{S\bar S}^{1/2}F^S\over\hat K_{T\bar T}^{1/2}F^T}=
\fc{1}{\sqrt{3}}\ 
e^{i(\alpha_T-\alpha_S)}\ {\int G_3\wedge \ov \Omega \over \int\ov G_3\wedge \ov\Omega}\ ,}
with \BIM
\eqn\goldstino{\eqalign{
\hat K_{S\bar S}^{1/2}\ F^S&=\sqrt 3\  C\  m_{3/2}\ e^{i\alpha_S}\ \sin\theta_g\ ,\cr
\hat K_{T\bar T}^{1/2}\ F^T&=\sqrt 3\  C\  m_{3/2}\ e^{i\alpha_T}\ \cos\theta_g\ ,}}
and $\alpha_S,\ \alpha_T$ are the phases of the respective $F$--terms. The
real constant $C$ follows from the relation 
$$\hat K_{S\bar S}\ |F^S|^2+\hat K_{T\bar T}\ |F^T|^2=
3\ C^2\ |m_{3/2}|^2=3\ |m_{3/2}|^2+\hat V\ ,$$
with:
\eqn\new{\eqalign{
|m_{3/2}|^2&=\fc{1}{3}\ \hat K_{T\bar T}\ |F^T|^2 \ ,\cr
\hat V&=\hat K_{S\bar S}\ |F^S|^2\ .}}
Hence, we have:
\eqn\CC{
C^2=1+{\hat V\over 3\,|m_{3/2}|^2}\ .}
We have $C=1$ for vanishing cosmological constant $\hat V$ and non--vanishing gravitino mass
$m_{3/2}$.
From {\it Eqs.} \boilK, \Fterms\ and \goldstino\ we obtain:
\eqn\Obtain{\eqalign{
\fc{\kappa_4^4\ \lambda^2}{(2\pi)^4\ \ap^2}\ \lf|\int_{X_6} G_3\wedge \Omega\ri|^2&=
2^3\ C^2\ |m_{3/2}|^2\ \cos^2\th_g\ (T-\ov T)^3\ (S-\ov S)\ ,\cr
\fc{\kappa_4^4\ \lambda^2}{(2\pi)^4\ \ap^2}\ \lf|\int_{X_6} \ov G_3\wedge \Omega\ri|^2&=
3\cdot 2^3\ C^2\ |m_{3/2}|^2\ \sin^2\th_g\ (T-\ov T)^3\ (S-\ov S)\ ,\cr
\fc{\kappa_4^4\ \lambda^2}{(2\pi)^4\ \ap^2}\ \int_{X_6} \ov G_3\wedge \Omega\ 
\int_{X_6} \ov G_3\wedge \ov \Omega&=
\sqrt 3\cdot 2^3\ C^2\ |m_{3/2}|^2\ \sin\th_g\ \cos\th_g\ e^{i\alpha_T-i\alpha_S}\ 
(T-\ov T)^3\ (S-\ov S)\ .}}
In the following, we shall parametrize the results of the last section with the goldstino angle.
\vskip0.2cm
\noindent
{\it (i) Untwisted scalar mass terms}

\noindent
Stack 1:
\eqn\umsa{\eqalign{
(m_{1\ov 1}^{7,1})^2&=C^2 |m_{3/2}|^2\ \kappa_4^2\ G_{C_1\ov C_1}^{7,1}\ 
\lf\{\ 3\ \lf(1+\fc{(T-\ov T)\ [18\ (S-\ov S)+T-\ov T]}{(S-\ov S)^2\ [9\ (S-\ov S)+T-\ov T]^2}
\ri)\ \sin^2\th_g\ri.\cr
&+\lf(1+\fc{9\ (S-\ov S)\ [9\ (S-\ov S)+2\ (T-\ov T)]}
{(T-\ov T)^2\ [9\ (S-\ov S)+T-\ov T]^2}\ri)\ \cos^2\th_g\cr
&\lf.-\fc{18 \sqrt 3}{[9\ (S-\ov S)+T-\ov T]^2}\  
\sin\th_g\ \cos\th_g\ \cos(\alpha_S-\alpha_T)\ \ri\},\cr
(m_{2\ov 2}^{7,1})^2&=3\ \kappa_4^2\ G_{C_2\ov C_2}^{7,1}\ C^2\ |m_{3/2}|^2\ \sin^2\th_g=
{3\over 2}\ {1\over |T-\ov T|}\ C^2\ |m_{3/2}|^2\ \sin^2\theta_g\ ,\cr
(m_{3\ov 3}^{7,1})^2&=3\ \kappa_4^2\ G_{C_3\ov C_3}^{7,1}\ C^2\ |m_{3/2}|^2\ \sin^2\th_g=
{3\over 2}\ {1\over |T-\ov T|}\ C^2\ |m_{3/2}|^2\ \sin^2\theta_g\ .}}
\noindent
Stack 2:
\eqn\umsb{\eqalign{
(m_{1\ov 1}^{7,2})^2&={3}\ \kappa_4^2\ C^2\ |m_{3/2}|^2\ \sin^2\theta_g\ G_{C^1\ov C^1}^{7,2}
={3\over 2}\ {1\over |T-\ov{T}|}\ C^2\ |m_{3/2}|^2\sin^2\theta_g\ ,\cr
(m_{2\ov 2}^{7,2})^2&=\kappa_4^2\ C^2\ |m_{3/2}|^2\ \cos^2\theta_g\ G_{C^2\ov C^ 2}^{7,2}
={1\over 4}\ C^2\ |m_{3/2}|^2\cos^2\theta_g\ ,\cr
(m_{3\ov 3}^{7,2})^2&={3}\ \kappa_4^2\ C^2\ |m_{3/2}|^2\ \sin^2\theta_g\ G_{C^3\ov C^3}^{7,2}
={3\over 2}\ {1\over |T-\ov{T}|}\ C^2\ |m_{3/2}|^2\sin^2\theta_g\ .}}
\vskip0.2cm
\noindent
{\it (ii) Twisted scalar mass terms}\br
For $T^i=T$, the twisted scalar masses \twistedas\ become ($a=3\sqrt\fc{S-\ov S}{T-\ov T}$):
\eqn\twistedas{\eqalign{
G_{C^{7_17_2}\ov C^{7_17_2}}&=2i\ \ \kappa_4^{-2}\  
(S-\ov S)^{-\fc{1}{4}+\fc{3}{2}\beta+\gamma}\ 
(T-\ov T)^{-\fc{3}{4}-\fc{3\beta}{2}-3\gamma}\cr
&\times \fc{1}{\pi}\ \sqrt{\fc{a}{1+a^2}}\ 
\Gamma\lf[\h-\fc{1}{\pi}\arctan(a)\ri]\ \Gamma\lf[\fc{1}{\pi}\arctan(a)\ri]\ ,\cr
G_{C^{7_27_3}\ov C^{7_27_3}}&=-2\ \kappa_4^{-2}\  (S-\ov S)^{-\fc{1}{4}+\fc{3}{2}\beta+\gamma}\ 
(T-\ov T)^{-\fc{3}{4}-\fc{3\beta}{2}-3\gamma}\cr
&\times\fc{1}{\pi}\ \sqrt{\fc{a}{1+a^2}}\ 
\Gamma\lf[\h+\fc{1}{\pi}\arctan(a)\ri]\ \Gamma\lf[-\fc{1}{\pi}\arctan(a)\ri]\ ,\cr
G_{C^{7_27_3}\ov C^{7_27_3}}&=-\kappa_4^{-2}\,(2i)^{-3/2}{1\over(T-\ov T)^{1/2}}\ .}}

\noindent
$1/4\ BPS\ \ ,\ \ b=2,3$:
\eqn\tmsq{\eqalign{
(m^{7_17_b})^2&=-\kappa_4^2\ C^2\ |m_{3/2}|^2\ \ 
\fc{G_{C^{7_17_b}\ov C^{7_17_b}}}{[9\ (S-\ov S)+T-\ov T]^2}\ \cr
&\times \lf\{\ 
3\ \lf(m_{SS}-[27\ (S-\ov S)+(T-\ov T)]\ \Psi^{(0)}_b(a)+\Psi^{(1)}_b(a)\ri)\ 
\sin^2\th_g\ri.\cr 
&
+\lf(m_{TT}+[9\ (S-\ov S)+3\ (T-\ov T)]\ \Psi^{(0)}_b(a)+\Psi^{(1)}_b(a)\ri)\ 
\cos^2\theta_g\cr
&\lf.+
2 \sqrt 3\  \lf(m_{ST}+[9\ (S-\ov S)-\ (T-\ov T)]\ \Psi^{(0)}_b(a)-\Psi^{(1)}_b(a)\ri)\ 
\sin\th_g\cos\th_g\cos(\alpha_S-\alpha_T)\ri\},}}
with
\eqn\Withdef{\eqalign{
m_{SS}&=-81\ (1+3\beta+2\gamma)\ (S-\ov S)^2-(2+3\beta+2\gamma)\ 
(T-\ov T)\ [18\ (S-\ov S)+(T-\ov T)]\ ,\cr
m_{TT}&=3\ (\beta+2\gamma)\ (T-\ov T)^2
-9\ (1-3\beta-6\gamma)\ (S-\ov S)\ [9\ (S-\ov S)+2\ (T-\ov T)]}}
and the expressions $\Psi^{(n)}_b,\ m_{ST}$ 
defined in \withdef, subject to the replacement $T^1\ra T$.

\noindent
$1/2\ BPS$:
\eqn\tmsh{\eqalign{
(m^{7_27_3})^2&=\h\ \ C^2\ \kappa_4^2\ 
|m_{3/2}|^2\ G_{C^{7_27_3}\ov C^{7_27_3}}\ (\ \cos^2\theta_g+3\,\sin^2\theta_g\ )\cr
&=2^{-5/2}\ \fc{1}{|T-\ov T|^{1/2}}\ C^2\ |m_{3/2}|^2\ (\ \cos^2\theta_g+3\,\sin^2\theta_g\ )\ .}}
\vskip0.2cm
\noindent
{\it (iii) Trilinear couplings}

\eqn\trilineara{\eqalign{
A_{ijk}^{7,1}&={1\over 4}\ \epsilon_{ijk}\ \fc{C\ m_{3/2}}{(T-\ov T)^{3/2}}
\ \left\{-{27\sqrt3\ (S-\ov S)
\over 9\ (S-\ov S)+(T-\ov T)}\ \sin\theta_g\ e^{i\alpha_s} \right.\cr&+\left.
\left[1+{27(S-\ov S)\over 9\ (S-\ov S)+(T-\ov T)}\right]\cos\theta_g\,
e^{i\alpha_T}\right\}\ ,\cr
A_{ijk}^{7,2}&=A_{ijk}^{7,3}={1\over 4}\ \epsilon_{ijk}\ \fc{C\ m_{3/2}} 
{(T-\ov T)^{3/2}}\ \cos\theta_g\ e^{i\alpha_T}\ .}}

\vskip0.2cm
\noindent
{\it (iv) Gaugino mass terms}
\eqn\gm{\eqalign{
m_{g,1}&=C\ m_{3/2}\ {(T-\ov T)\ \cos\theta_g\ e^{i\alpha_T}+9(S-\ov S)\sqrt3\ \sin\theta_g\ 
e^{i\alpha_S}\ \over(T-\ov T)+9(S-\ov S)}\ ,\cr
m_{g,2}&=m_{g,3}=C\ m_{3/2}\ \cos\theta_g\ e^{i\alpha_T}\ .}}

\newsec{Gravitino mass and scales: scalar and gaugino masses}

In this section, we shall discuss the scales of the soft--supersymmetry breaking terms.
The key quantity entering all formulae is the product $\Cc:=C\ m_{3/2}$, with $C$ 
introduced in \eqq \CC\ and the gravitino mass $m_{3/2}=|\hat W|\ \kappa_4^2\ 
e^{\h\kappa_4^2\ \hat K}$. Here, the K\"ahler potential $\hat K$ is given in \eqq \boilK, 
and the superpotential $\hat W$ in \eqq \TVW.
In the general case of turning on both $(3,0)$-- and $(0,3)$--flux components, \ie
generic goldstino angle $\th_g\neq 0,\fc{\pi}{2},\ldots$ we have
\eqn\general{
\Cc=\sqrt{m_{3/2}^2+\fc{1}{3}\ \hat V}\ ,}
which boils down to
\eqn\boil{
\Cc=\cases{   m_{3/2}\ ,                            & $(0,3)$--flux\ ,\cr
                   \fc{1}{\sqrt 3}\ \hat V^{1/2}\ , & $(3,0)$--flux}}
for the two special cases $\theta_g=0$ and $\theta_g=\fc{\pi}{2}$, respectively.
Recall, that only for $IASD$--fluxes the 
cosmological constant $\hat V$ is non--vanishing (to lowest order), while only a $(0,3)$--flux 
component $G_3$ gives rise to a non--vanishing gravitino mass $m_{3/2}$.
Hence, in either case the quantity $\Cc$ is non--vanishing and generically allows for 
non--vanishing soft--masses in the following.
The  gravitino mass may be written
\eqn\gravmass{\eqalign{
m_{3/2}&=\fc{1}{\sqrt 2\ (2\pi)^6}\ \ \fc{M_{\rm string}^8}{M_{\rm Planck}^2}\ \ 
\fc{1}{\im(S)^{1/2}\ \prod\limits_{j=1}^3\im(T^j)^{1/2}\im(U^j)^{1/2}}\ \ 
\lf|\int_{X_6} G_3\wedge \Omega\ \ri|\cr
&=\fc{\gs^2}{\sqrt 2\ (2\pi)^4}\ \ \fc{M_{\rm string}^2}{M_{\rm Planck}^2}\ \   
\fc{\prod\limits_{j=1}^3\im(U^j)^{-1/2}}{{\rm Vol}(X_6)}\ \ 
\lf|\int_{X_6} G_3\wedge \Omega\ \ri|\ ,}}
with the \tb string coupling constant $g_{\rm string}=e^{\phi_{10}}=(2\pi\ \im S)^{-1}$.
The latter is assumed to be small in order to justify a perturbative orientifold construction.
The factor ${\rm Vol}(X_6)=\im(\Tc^1)\im(\Tc^2)\im(\Tc^3)$ is the volume\foot{The following
relations have been used: 
$\phi_4=\phi_{10}-\h\ln[\im(\Tc^1)\im(\Tc^2) \im(\Tc^3)/\ap^3]$,\ $\kappa_{10}^{-2}=
\fc{2}{(2\pi)^7}\ \ap^{-4}$, and $e^{\kappa_4^2\hat K}=
\fc{(2\pi)^4\ e^{4\phi_4}}{2^7\ \prod\limits_{j=1}^3 \im U^j}$. 
Moreover, we have: $\fc{\im(S)^3}{
\prod\limits_{j=1}^3 \im(T^j)}=\fc{\ap^6}{{\rm Vol}(X_6)^2}$. Consult {\it Ref.} \LMRS\ for more 
details. Besides, in \TVW\ we have choosen 
$\lambda^{-1}=16\pi^5\ap^3$, such that $\kappa_{10}^{-2}=\fc{\lambda}{(2\pi)^2\ap}$.} 
of the six--dimensional compactification manifold $X_6$, measured in string units $\ap^3$.
The relation between the string scale $\ap=M_{\rm string}^{-2}$ and the four--dimensional 
Planck mass $M_{\rm Planck}$ is given by: 
\eqn\Planck{
M_{\rm Planck}=2^{3/2}\ \ g_{\rm string}^{-1}\ M_{\rm string}^4\ \sqrt{{\rm Vol}(X_6)}\ .}
Qualitatively, the integral $|\int_{X_6} G_3\wedge \Omega|$ is of
order $\fc{M_{\rm Planck}}{M_{\rm string}^6}$. Since the moduli fields $S,T^j,U^j$ are 
dimensionsless we deduce from the first line of \gravmass: 
$m_{3/2}\sim\fc{M_{\rm string}^2}{M_{\rm Planck}}$.

In the following, as in the previous section,  
let us assume an isotropic compactification of radius $R$, \ie
${\rm Vol}(X_6)=R^6$ and $U^j=i$.
The latter clearly obeys the supersymmetry condition \susya.
The flux quantization condition 
$\fc{1}{(2\pi)^2\ap}\ \int_{C_3}G^{(0,3)}_3=\xi_1\in \IZ$ for a $(0,3)$--form flux component
of $G_3$ essentially yields the estimate:
\eqn\fluxq{
G_3^{(0,3)} \sim (2\pi)^2\ \fc{\xi_1\ \ap}{R^3}\ .}
With this information and
$$\lf|\ \int_{X_6}G_3\wedge\Omega\ \ri|=(2\pi)^8\ \xi_1\ \ap\ R^3=2^{-3/2}\ (2\pi)^8\ \gs\ 
\fc{M_{\rm Planck}}{M_{\rm string}^6}\ \ \xi_1\ ,$$
we obtain for the gravitino mass $m_{3/2}$:
\eqn\obtain{
m_{3/2}=\pi^2\ \ \fc{1}{\im(S)^{1/2}\im(T)^{3/2}}\ \ \fc{M_{\rm string}^2}{M_{\rm Planck}}
\ \ \xi_1\ .}
Since the physical moduli fields $T$ are dimensionless, we have:
\eqn\WEhave{
m_{3/2}\ \sim\ \gs^{1/2}\ \ \fc{M_{\rm string}^2}{M_{\rm Planck}}\ \ \xi_1\ .}
Hence, for the model introduced in subsection 3.3, with $\im(S)=1$ as a result from the
flux quantization condition, we obtain:
\eqn\obtainfinal{
m_{3/2}=\pi^2\ \ \im(T)^{-3/2}\ \ 
M_{\rm string}\ \lf(\fc{M_{\rm string}}{M_{\rm Planck}}\ri)\ \xi_1\ .}
To relate the flux density $\xi_1$ to the goldstino angle and the quantity $C$, 
introduced in the previous section, we also define the density $\xi_2$ for a 
pure $(3,0)$--flux component: 
$\fc{1}{(2\pi)^2\ap}\ \int_{C_3}G^{(3,0)}_3=\xi_2\in \IZ$, \ie 
$G_3^{(3,0)} \sim (2\pi)^2\ \fc{\xi_2\ \ap}{R^3}$.
Then, we obtain the following relations:
\eqn\Relations{\eqalign{
\fc{\xi_1}{\xi_2}&=\fc{m_{3/2}}{\sqrt{\hat V}}\ ,\cr
\xi_2&=\xi_1\ \sqrt3\ \tan\th_g\ .}}
Besides, we have:
\eqn\besides{
\xi_2=\pi\ \lf(\fc{M_{\rm Planck}}{M_{\rm string}}\ri)^3\ M_{\rm string}^{-1}\ \sqrt{\hat V}\ .}
Hence, in the following, whenever only the density $\xi_1$ appears, the relations \Relations\
allow us to replace $\xi_1$ by $\xi_2$ through the goldstino angle.

{From} the relations \WEhave\ or \obtainfinal\ 
we see, that for $M_{\rm string}\sim M_{\rm Planck}$, the flux density
$\xi_1$ has to be very small, in order to arrive at a small gravitino mass. In other words,
the flux has to be largely thinned out over space--time. The latter effect may be 
achieved with a large warping suppression. On the other hand, if 
the string scale $M_{\rm string}$ is sufficiently low (\eg $M_{\rm string}\sim 10^{11}\ GeV$ and
$R\sim 10^{-9}\ GeV^{-1}$), 
reasonable values for $\xi_1$ may be chosen.
Similar conclusions apply for the soft masses, which have been derived in the previous 
two sections. We shall discuss their scales in the following.

With the assumption of an isotropic 
compactification\foot{The closed string moduli fields, introduced in \eqqs \fieldST, may be 
written in terms of $g_{\rm string}$ as:
\eqn\moduli{\eqalign{
T^j&=a^j+i\ \fc{\gs^{-1}}{2\pi\ap^2}\ \im\Tc^k\im\Tc^l\ ,\cr
S&=C_0+i\ \fc{\gs^{-1}}{2\pi}\ .}}
Hence, for an isotropic compactification, which respects the supersymmetry constraint \susya, 
we have $\im\tilde T=\im T^j=\fc{\gs^{-1}}{2\pi\ap^2}\ R^4$.} we obtain\foot{
For the masses of the fields $C^{7_17_2}, C^{7_17_3}$, given in \tmsq, 
we shall only show their power series w.r.t. $M_{\rm string}/{M_{\rm Planck}}$.}  
for the (untwisted sector) scalar masses
\umsa\ and \umsb:
\eqn\Umsb{\eqalign{
m_{2\ov 2}^{7,1}=m_{3\ov 3}^{7,1}=m_{1\ov 1}^{7,2}=m_{3\ov 3}^{7,2}
&=\lf(\fc{3\pi}{2}\ri)^{1/2}\ \gs^{1/2}\ \fc{\ap}{R^2}\ \Cc\ \sin\theta_g\cr
&=\sqrt{6\pi}\ \gs^{-1/6}\ \lf(\fc{M_{\rm string}}{M_{\rm Planck}}\ri)^{2/3}\ 
\Cc\ \sin\theta_g\ ,\cr
m_{2\ov 2}^{7,2}=m_{3\ov 3}^{7,3}&=\h\ \Cc\ \cos\theta_g\ ,\cr
m_{1\ov 1}^{7,1}&=2^{-5/2}\ \Cc\ \lf(1+9\ \fc{\ap^2}{R^4}\ri)^{-1/2}\cr
&\hskip-5cm\times
\lf\{\ 13+234\ \fc{\ap^2}{R^4}-18\ \fc{\ap^6}{R^{12}}-81\ \fc{\ap^8}{R^{16}}-
\lf(5+90\ \fc{\ap^2}{R^4}+648\ \fc{\ap^4}{R^8}+18\ \fc{\ap^6}{R^{12}}+81\ \fc{\ap^8}{R^{16}}\ri)\ 
\cos(2\th_g)\ri.\cr
&\lf.+18\ \fc{\ap^4}{R^8}\ [\ 72+\sqrt 3\ \sin(2\th_g)\ ]\ \ri\}^{1/2}\ .}}
Furthermore, the soft mass for the ($1/2$ BPS) twisted matter field $C^{7_27_3}$ takes the form:
\eqn\Tmsh{\eqalign{
m^{7_27_3}&=2^{-5/4}\ \pi^{1/4}\ \gs^{1/4}\ \fc{\ap^{1/2}}{R}\ \Cc\ 
\sqrt{2-\cos(2\th_g)}\cr
&=2^{-3/4}\ \pi^{1/4}\ \gs^{-1/12}\ \lf(\fc{M_{\rm string}}{M_{\rm Planck}}\ri)^{1/3}\ \Cc
\ \sqrt{2-\cos(2\th_g)}\ .}}
{From} the above equations \Umsb\ and \Tmsh, we deduce that 
the soft--masses for the untwisted matter fields and $m^{7_27_3}$ are roughly of 
the same order $\Oc(m_{3/2})$ 
for the case of $M_{\rm string}\sim M_{\rm Planck}$, $\gs\sim (2\pi)^{-1}$ and
a goldstino angle $\theta_g\neq 0,\fc{\pi}{2},\ldots$. 
The masses $m_{2\ov 2}^{7,1},\ m_{3\ov 3}^{7,1},\ m_{1\ov 1}^{7,2},\ m_{3\ov 3}^{7,2}$
of the Wilson line  moduli from the $D7$--branes without $2$--form fluxes stay massless
in the case of a pure $(0,3)$--form flux, \ie $\th_g\sim 0$.
Contrarily, the moduli $C^{7,2}_2,\ C^{7,3}_3$ 
describing the positions of the second and third stack of $D7$--branes become
massive in the case of a $(0,3)$--form flux. Furthermore, the modulus $C^{7,1}_1$ 
describing the position of the first stack of $D7$--branes 
with non--vanishing $2$--form flux becomes massive through the combined
effect of a $(0,3)$--form flux and the $2$--form flux. The mixing with an additional 
$(3,0)$--form flux is described by the goldstino angle.
There is a non--universality in the masses $m_{2\ov 2}^{7,2},\ m_{3\ov 3}^{7,3}$
and $m_{1\ov 1}^{7,1}$. This effect is increased by the goldstino angle.
However, as we shall see in a moment, 
this universality disappears for string scales $M_{\rm string}\ll M_{\rm Planck}$.

In the following, the cosmological constant $\hat V$ is assumed to be small.
According to \general, we may choose $\Cc\sim m_{3/2}$ in the above equations. Moreover, we 
expand the soft--masses w.r.t. the ratio $\fc{M_{\rm string}}{M_{\rm Planck}}\ll 1$
(\cf the discussion above).
This leads to the following estimates for the scalar masses of the untwisted sector:
\eqn\UUmsb{\eqalign{
m_{2\ov 2}^{7,1}=m_{3\ov 3}^{7,1}=m_{1\ov 1}^{7,2}=m_{3\ov 3}^{7,2}
&=(2\pi)^{2/3}\ \sqrt 3\  
m_{3/2}\ \lf(\fc{M_{\rm string}}{M_{\rm Planck}}\ri)^{2/3}\ \sin\theta_g\ ,\cr
m_{2\ov 2}^{7,2}=m_{3\ov 3}^{7,3}&=\h\ m_{3/2}\ \cos\th_g\ ,\cr
m_{1\ov 1}^{7,1}&=2^{-5/2}\ m_{3/2}\ \sqrt{13-5\ \cos(2\th_g)}\ \ .}}
In order to keep the cosmological constant small, one should aim for a small 
goldstino angle $\th_g$, in which case supersymmetry breaking is mainly due to  
$(0,3)$--flux components. 
In the regime of small ratio $\fc{M_{\rm string}}{M_{\rm Planck}}\ll 1$,
we observe a universality in the untwisted sector masses: The scalar masses
$m_{2\ov 2}^{7,1},\ m_{3\ov 3}^{7,1},\ m_{1\ov 1}^{7,2},\ m_{3\ov 3}^{7,2}$
referring to the Wilson line moduli vanish for a small goldstino angle, while
the masses $m_{1\ov 1}^{7,1}, m_{2\ov 2}^{7,2}, m_{3\ov 3}^{7,3}$ 
referring to the $D7$--brane positions become equal for $\th_g=0$.
On the other hand, the latter vanish for $\th_g \sim \fc{\pi}{2}$.
Of course, this observation is just in lines with the fact that a pure $(3,0)$--form flux
gives rise to the scalar masses of the Wilson line moduli only, while a pure $(0,3)$--form
flux gives masses only to the transverse $D7$--brane position moduli. 
Note, that the $2$--form flux dependence of $m_{1\ov 1}^{7,1}$  has completely disappeared in the
limit  $\fc{M_{\rm string}}{M_{\rm Planck}}\ll 1$. Hence, the universality is independent
of the $2$--form flux turned on, at least to lowest order in $\fc{M_{\rm string}}{M_{\rm Planck}}$.

Let us now turn to the expansion of the scalar masses of the twisted sector, given in \eqqs
\tmsh\ and \tmsq:
\eqn\TTmsh{\eqalign{
m^{7_27_3}&= 2^{-2/3}\ \pi^{1/3}\ m_{3/2}\ \lf(\fc{M_{\rm string}}{M_{\rm Planck}}\ri)^{1/3}\ 
\sqrt{2-\cos(2\th_g)}\ ,\cr
m^{7_17_2}=m^{7_17_3}&\sim 3^{-1/4}\ 2^{\fc{1}{3}+\fc{5}{2}\beta+4\gamma}\ 
\pi^{\fc{7}{12}+\beta+2\gamma}\ m_{3/2}\ 
\lf(\fc{M_{\rm string}}{M_{\rm Planck}}\ri)^{\fc{1}{3}+\beta+2\gamma}\cr
&\times \sqrt{2+3\beta-(1+6\beta+6\gamma)\ \cos(2\th_g)}\ .}}
Let us compare the $1/4$ BPS sector masses $m^{7_17_2}, m^{7_17_3}$ with the Higgs mass 
$m^{7_27_3}$. We obtain  the following ratio
\eqn\behaviour{\eqalign{
\fc{m^{7_17_2}}{m^{7_27_3}}=\fc{m^{7_17_3}}{m^{7_27_3}}&=3^{-1/4}\ 2^{1+\fc{5}{2}\beta+4\gamma}\ 
\pi^{\fc{1}{4}+\beta+2\gamma}\ \lf(\fc{M_{\rm string}}{M_{\rm Planck}}\ri)^{\beta+2\gamma}\cr
&\times \sqrt{\fc{2+3\beta-(1+6\beta+6\gamma)\ \cos(2\th_g)}{2-\cos(2\th_g)}}\ ,}}
which becomes
\eqn\becomes{
\fc{m^{7_17_2}}{m^{7_27_3}}=\fc{m^{7_17_3}}{m^{7_27_3}}=3^{-1/4}\ 2^{1+\fc{5}{2}\beta+4\gamma}\ 
\pi^{\fc{1}{4}+\beta+2\gamma}\ \lf(\fc{M_{\rm string}}{M_{\rm Planck}}\ri)^{\beta+2\gamma}
\ \sqrt{1-3\beta-6\gamma}}
for $\th_g\ra 0$. 
Hence, the ratio is very sensitive to the constants $\beta,\gamma$ to be determined in \wip. 
For $\beta,\gamma\neq 0$ the ratios \becomes\ are in lines of 
a split--SUSY scenario \lref\ArkaniHamedFB{
  N.~Arkani-Hamed and S.~Dimopoulos,
  ``Supersymmetric unification without low energy supersymmetry and signatures
  for fine-tuning at the LHC,''
  arXiv:hep-th/0405159.
} \ArkaniHamedFB.

Finally, let us discuss the gaugino masses, which have been presented in \eqq \gm:
\eqn\GM{\eqalign{
m_{g,2}=m_{g,3}&= \Cc\ \cos\th_g\ e^{i\alpha_T}\ ,\cr
m_{g,1}&=\Cc\ \lf(1+9\ \fc{\ap^2}{R^4}\ri)^{-1}\ \lf(e^{i\alpha_T}\ \cos\theta_g+
9\sqrt3\ e^{i\alpha_S}\ \fc{\ap^2}{R^4}\ \sin\theta_g\ri)\ .}}
As already anticipated in the introduction, the gaugino masses referring to 
$D7$--brane stacks without $2$--form fluxes are only sensitive to $(0,3)$--flux components,
\ie their masses are proportional to $\cos\theta_g$. On the other hand, $D7$--branes with
non--vanishing $2$--form fluxes on their internal world--volume lead to gaugino masses, which
feel both $(0,3)$-- and $(3,0)$--flux components. In other words, $m_{g,1}$ is generically 
non--vanishing. However, the dependence of $m_{g,1}$ on the $(3,0)$--flux component
is sub--leading in $\ap.$ In particular, all three stacks give rise to gaugino masses
with the same leading power behaviour w.r.t. $M_{\rm string}/M_{\rm Planck}$:
\eqn\GGM{
m_{g,1}=m_{g,2}=m_{g,3}=m_{3/2}\ \cos\theta_g\ e^{i\alpha_T}\ .}
Hence, like the scalar masses the gaugino masses are universal for 
$\fc{M_{\rm string}}{M_{\rm Planck}}\ll 1$.

\newsec{Concluding remarks}

In this paper we have computed the $G$--flux induced soft supersymmetry
breaking terms in semirealistic $D$--brane models, in which the
gauge/matter sector of the MSSM originated from
open strings on $D7$--branes with $f$--flux.
Specifically, the matter fields, namely quarks, leptons, Higgs fields and
their N=1 superpartners correspond to twisted open string
sectors ending on $D7$--branes
with different $f$--flux boundary conditions.
The analysis was performed in a local $D7$--brane
model whose
twisted spectrum is just the one of the MSSM.
Tadpole cancellation on 
a compact orbifold will require additional hidden
sector $D7$--branes with $f$--flux.
The soft masses were computed as a function
of the K\"ahler moduli, being still partially 
unfixed despite the supersymmetry conditions,
and also as a function of two
3--form flux components, the $(0,3)$--flux $G_{(0,3)}$ and
the $(3,0)$--flux   $G_{(3,0)}$.
In order to obtain a tiny cosmological constant together
with a non-vanishing gravitino mass,
the  $G_{(3,0)}$ flux component must be much smaller compared to
$G_{(0,3)}$. In other words, the goldstino angle $\theta_g$, 
introduced in subsection 3.4, has to be very small. 
Moreover, in order to keep $m_{3/2}$ much below
$M_{\rm Planck}$, namely in the $TeV$--region,
the string scale must be sufficiently low.
Then the gaugino masses are also of the order of $m_{3/2}$ 
(\cf section 4).
The squark and slepton masses exhibit
a more complicated moduli dependence and are in general non-universal,
as they depend on the $f$--fluxes of the involved $D7$--branes
(the intersection angles in \ta language with
intersecting $D6$--branes). As a result of our analysis in section 5,
it turns out that for a low string scale, the squark and slepton masses are
considerably different than the SUSY--breaking 
mass contribution to the Higgs field (\cf the ratio \becomes). This may be in favor
for a split--SUSY scenario \ArkaniHamedFB.
The soft masses of squark and slepton
fields in different families, \ie open
strings sitting at different `intersection angles',
are the same, as long as the `intersection angles' 
of the different families agree.
This will usually be the case in concrete MSSM--like
models. Finally, the gaugino masses are typically of the same order as $m_{3/2}$
(\cf \eqq \GGM).

The orientifold models on orbifold backgrounds considered here, so far
suffer at least one serious phenomenological problem:
as we discussed, there will be also matter fields
from untwisted open string sectors (N=4 sectors)
with open strings ending on the same $D7$--branes.
The corresponding scalar and fermion fields
transform in the adjoint representation of
the gauge group factors, and they correspond to
Wilson line fields or scalars,
that describe the locations of the $D7$--branes, commonly denoted by $D7$--brane moduli.
Without any other 3-form fluxes as 
$G_{(0,3)}$ and $G_{(3,0)}$ turned on, the adjoint fermions will stay massless
and the adjoint scalar components will get a soft mass,
shown in \eqqs \Umsb\ and \UUmsb. 
In the MSSM sector, this is clearly unacceptable since
these states are not observed. But also
in the hidden sectors they cause a serious
problem, because they lead to a negative (non-asymtotically free)
$\beta$-function, which forbids a non-perturbative superponential,
e.g. by gaugino condensation.
Therefore, additional effects are required in order
to give these adjoint multiplets a large
supersymmetry preserving mass.

One possibility to acquire this kind of
desired mass terms is 
to consider $F$--theory with $4$--form flux $G_4$
turned on, which gives rise to a flux superpotential of the 
form $W\sim \int G_4\wedge \Omega_4$. Here, the $D7$--brane moduli represent 
complex structure moduli of the fourfold and therefore naturally enter the superpotential.
A concrete example of this type, namely $F$--theory
on $K3\times K3$, was recently discussed
in {\it Ref.} \GorlichQM, where a flux
induced supersymmetric mass term for the $D7$--brane moduli
was indeed generated. The latter result 
agrees with computations of gauged supergravity \ADFT.
However, the \tb interpretation of this
mass term is still somewhat unclear, since
in the \tb language, a mass for the open string
moduli has to be generated, whereas the $3$--form flux
induced superpotential \TVW\ a priori
only depends on the closed string moduli fields.
Nevertheless, this mass term for the $D7$--brane
moduli will arise in \tb orientifolds
with N=2 sectors after turning on a suitable supersymmetric
$(2,1)$--flux component, as we will show in {\it Ref.} \LRSm.

\bigskip
\centerline{\bf Acknowledgments }\nobreak
\bigskip
D.L. and St.St.  
thank the Humboldt University in Berlin for hospitality. S.R. thanks the University 
of Munich for hospitality.
We would like to thank R.~Blumenhagen for helpful discussions.
This work is supported in part by the Deutsche 
Forschungsgemeinschaft (DFG), and the German--Israeli Foundation (GIF).

\appendix\appA{Curvature tensors for untwisted and twisted $D7$--brane fields}

For the untwisted matter fields of the stacks without $f$-flux, the curvature
tensors take a particularly simple form. We get only the following non-zero 
elements:
\eqn\Runnof{\eqalign{
R^{7,2}_{U^i\ov U^i i\ov i}&={-1\over (U^i-\ov U^i)^2}\ G^{7,2}_{C_i\ov C_i}\ ,\ 
i=1,2,3\ ,\cr
R^{7,2}_{T^3\ov T^3 1\ov 1}&={-1\over (T^3-\ov T^3)^2}\ G^{7,2}_{C_1\ov C_1}\ ,\cr
R^{7,2}_{S\ov S 2\ov 2}&={-1\over (S-\ov S)^2}\ G^{7,2}_{C_2\ov C_2}\ ,\cr
R^{7,2}_{T^1\ov T^1 3\ov 3}&={-1\over (T^1-\ov T^1)^2}\ G^{7,2}_{C_3\ov C_3}\ .}}
For stack 3, we get the same result with indices 2 and 3 interchanged.

{From} now on, we will take $M,N$ to run over $S,\,T^i$ only as we consider 
here the case with $(3,0)$-- and $(0,3)$--fluxes only.

Due to the non-vanishing $f$-flux on stack 1, the form of the curvature tensor
is here much more involved, the components with mixed moduli are no longer
zero and the expression is long and ugly, the reason for which being the
dependence of the ${\cal T}^i$ on all the $T^i$ and on $S$:
$$\eqalign{
\ti&=\ap\ \lf(\fc{T_2^2\ T_2^3}{S_2\ T_2^1}\ri)^{1/2}\ \ \ ,\ \ \ 
\tii=\ap\ \lf(\fc{T_2^1\ T_2^3}{S_2\ T_2^2}\ri)^{1/2}\ ,\cr
\tiii&=\ap\ \lf(\fc{T_2^1\ T^2_2}{S_2\ T_2^3}\ri)^{1/2}\ \ \ ,\ \ \ 
e^{-\phi_4}=2\pi\ (S_2\ T_2^1\ T_2^2\ T_2^3)^{1/4}\ .}
$$
Here, we need to know the $\partial_M({\rm Im}{\cal T}^j)$:
\eqn\deriv{\eqalign{
{\partial({\rm Im}{\cal T}^j)\over \partial S}&={i\over4}{{\rm Im}{\cal T}^j\over
{\rm Im}S}\ ,\cr
{\partial({\rm Im}{\cal T}^j)\over \partial T^j}&={i\over 4}{{\rm Im}{\cal T}^j\over
{\rm Im}T^j}\ ,\cr
{\partial({\rm Im}{\cal T}^j)\over \partial T^k}&={-i\over 4}{{\rm Im}{\cal T}^j\over
{\rm Im}T^k}\ \ \ ,\ \ \ j\neq k\ .}}
We will also need the $\partial_M\partial_{\ov N}({\rm Im}{\cal T}^j)$:
\eqn\doublederiv{\eqalign{
{\partial({\rm Im}{\cal T}^j)\over {\partial S \partial \ov S}}&=
{3\over 16}{({\rm Im}{\cal T}^j)\over({\rm Im}S)^2}, \qquad
{\partial({\rm Im}{\cal T}^j)\over {\partial T^j \partial \ov S}}=
{1\over 16}{({\rm Im}{\cal T}^j)\over({\rm Im}S)({\rm Im}T^j)}\ ,\cr
{\partial({\rm Im}{\cal T}^j)\over {\partial T^k \partial \ov S}}&=
{-1\over 16}{({\rm Im}{\cal T}^j)\over({\rm Im}S)({\rm Im}T^k)},\quad
{\partial({\rm Im}{\cal T}^j)\over {\partial T^k \partial \ov T^i}}=
{1\over 16}{({\rm Im}{\cal T}^j)\over({\rm Im}T^i)({\rm Im}T^k)}\ ,\cr
{\partial({\rm Im}{\cal T}^j)\over {\partial T^k \partial \ov T^j}}&=
{-1\over 16}{({\rm Im}{\cal T}^j)\over({\rm Im}T^j)({\rm Im}T^k)},\quad
{\partial({\rm Im}{\cal T}^j)\over {\partial T^k \partial \ov T^k}}=
{-1\over 16}{({\rm Im}{\cal T}^j)\over({\rm Im}T^k)^2}\ ,\cr
{\partial({\rm Im}{\cal T}^j)\over {\partial T^j \partial \ov T^j}}&=
{3\over 16}{({\rm Im}{\cal T}^j)\over({\rm Im}T^j)^2}\ \ \ ,\ \ \ i\neq j\neq k\ ,}}
where $M,\ N$ run over $S,\ T^i$.

After some algebra we get the following results:
\eqn\Rmixed{\eqalign{
R^{7,1}_{M\ov N 2\ov 2}&={G^{7,1}_{C_2\ov C_2}\over \im M\ \im N}\ 
{(g\alpha')^2\over[(g\alpha')^2+(\im\Tc^2)^2]^2[(g\alpha')^2+(\im\Tc^3)^2]^2}\cr
&\times\lf\{(g\alpha')^6\ (\ \beta_2(M,N)-\beta_3(M,N)-\alpha_2(M,N)+\alpha_3(M,N)\ )\ri.\cr
&+(g\alpha')^4\ [\ (\ \beta_2(M,N)-2\beta_3(M,N)-3\alpha_2(M,N)+2\alpha_3(M,N)\ )\ (\im\Tc^2)^2\cr
&+(\ 2\beta_2(M,N)-\beta_3(M,N)-2\alpha_2(M,N)+3\alpha_3(M,N)\ )\ (\im\Tc^3)^2\ ]\cr
&+(g\alpha')^2\ [(-\beta_3(M,N)+\alpha_3(M,N))\ (\im\Tc^2)^4+(\beta_2(M,N)-\alpha_2(M,N))\ 
(\im\Tc^3)^4\cr
&+2\ (\ \beta_2(M,N)-\beta_3(M,N)-3\alpha_2(M,N)+3\alpha_3(M,N)\ )(\im\Tc^2)^2(\im\Tc^3)^2\ ]\cr
&+(\im\Tc^2)^2\ (\im\Tc^3)^2[\ -(\beta_3(M,N)-3\alpha_3(M,N))(\im\Tc^2)^2\cr
&\lf.+(\ \beta_2(M,N)-3\alpha_2(M,N)\ )\ (\im\Tc^3)^2\ ]\ri\}\ \ ,\ \  M, N\neq T^3\ ,\cr
R^{7,1}_{T^3 \ov T^3 2\ov{2}}&=\fc{G^{7,1}_{C_2\ov C_2}}{4\ (\im T^3)^2}\ 
\left\{1-{1\over 2}{(g\alpha')^2\ [(g\alpha')^2+2(\im\Tc^2)^2]\over[(g\alpha')^2+(\im\Tc^2)^2]^2}
-{1\over 2}{(g\alpha')^4\over[(g\alpha')^2+(\im\Tc^3)^2]^2}\ri\}\ ,}}
\eqn\Rmixedv{\eqalign{
R^{7,1}_{M\ov N1\ov{1}}&={-(g\alpha')^2\ G^{7,1}_{C_1\ov
C_1}\over[(g\alpha')^2+\im\Tc^2\im\Tc^3]^2}{1\over \im M\im
N}\left\{-[(g\alpha')^2+2\im\Tc^2\im\Tc^3]\ [\alpha_3(M,N)+\alpha_2(M,N)]\right.\cr
&+[(g\alpha')^2+\im\Tc^2\im\Tc^3]\ [\beta_3(M,N)+\beta_2(M,N)]\cr
&\left.-\im\Tc^2\im\Tc^3[\gamma_3(M)\ov\gamma_2(N)+\gamma_2(M)\ov\gamma_3(N)]
\right\}\ \ ,\ \ M, N\neq S\ ,\cr 
R^{7,1}_{S\ov S1\ov{1}}&={G^{7,1}_{C_1\ov C_1}\over 4\ (\im S)^2}\ \left\{1-
{(g\alpha')^4\over[(g\alpha')^2+\im\Tc^2\im\Tc^3]^2}\right\}\ .}}
The constants $\alpha_{2,3},\ \beta_{2,3}$ and $\gamma_{2,3}$ are summarized
in table A.1 and A.2.

Now we will look at the twisted case. 
The curvature tensor for the sector $(23)$ must be treated separately from the
one for $(12)$, $(13)$, as these sectors do not possess the same amount of
supersymmetry and have different metrics. We first examine the $1/4\ BPS$ sector.
The non-zero components of the curvature are:
\eqn\twistedseven{\eqalign{
R^{7a,7b}_{S\ov S}&=G_{C^{7a7b} \ov C^{7a7b}}\ \lf\{\fc{-1+6\beta+4\gamma}{4\ (S-\ov S)^2}
+\h\sum_{l=1}^3 \p_S\th^l_{ab}\ \p_{\ov S}\th^l_{ab}\ 
\lf[\ \psi_1(\th_{ab}^l)-\psi_1(1-\th^l_{ab})\ \ri]\ri.\cr
&+\lf.\sum_{l=1}^3\p_S\p_{\ov S} \th_{ab}^l \lf(\gamma\ \ln(T^l-\ov T^l)-\ln(U^l-\ov U^l)+
\h\ \lf[\ \psi(1-\th^l_{ab})+\psi(\th_{ab}^l)\ \ri]\ri)\ri\}\ ,\cr
R^{7a,7b}_{S\ov T^j}&=G_{C^{7a7b} \ov C^{7a7b}}\ \lf\{\fc{-\gamma\ \p_S\th^j_{ab}}{(T^j-\ov T^j)}
+\h\sum_{l=1}^3 \p_S\th^l_{ab}\ \p_{\ov T^j}\th^l_{ab}\ 
\lf[\ \psi_1(\th_{ab}^l)-\psi_1(1-\th^l_{ab})\ \ri]\ri.\cr
&+\lf.\sum_{l=1}^3\p_S\p_{\ov T^j} \th_{ab}^l \lf(\gamma\ \ln(T^l-\ov T^l)-\ln(U^l-\ov U^l)+
\h\ \lf[\ \psi(1-\th^l_{ab})+\psi(\th_{ab}^l)\ \ri]\ri)\ri\}\ ,\cr
R^{7a,7b}_{T^j\ov T^j}&=G_{C^{7a7b} \ov C^{7a7b}}\ \lf\{-\fc{1+2\beta+4\gamma(1-\th_{ab}^j)}
{4\ (T^j-\ov T^j)^2}-\fc{\gamma\ (\p_{T^j}\th_{ab}^j-\p_{\ov T^j}\th_{ab}^j)}{T^j-\ov T^j}\ri.\cr
&+\h\sum_{l=1}^3 \p_{T^j}\th^l_{ab}\ \p_{\ov T^j}\th^l_{ab}\ 
\lf[\ \psi_1(\th_{ab}^l)-\psi_1(1-\th^l_{ab})\ \ri]\cr
&+\lf.\sum_{l=1}^3\p_{T^j}\p_{\ov T^j} \th_{ab}^l \lf(\gamma\ \ln(T^l-\ov T^l)-\ln(U^l-\ov U^l)+
\h\ \lf[\ \psi(1-\th^l_{ab})+\psi(\th_{ab}^l)\ \ri]\ri)\ri\}\ ,\cr
R^{7a,7b}_{T^i\ov T^j}&=G_{C^{7a7b} \ov C^{7a7b}}\ \lf\{-\fc{\gamma\ \p_{T^i}\th^j_{ab}}{(T^j-\ov T^j)}
+\fc{\gamma\ \p_{\ov T^j}\th^i_{ab}}{(T^i-\ov T^i)}\ri.\cr
&+\h\sum_{l=1}^3 \p_{T^i}\th^l_{ab}\ \p_{\ov T^j}\th^l_{ab}\ 
\lf[\ \psi_1(\th_{ab}^l)-\psi_1(1-\th^l_{ab})\ \ri]\cr
&+\lf.\sum_{l=1}^3\p_{T^i}\p_{\ov T^j} \th_{ab}^l \lf(\gamma\ \ln(T^l-\ov T^l)-\ln(U^l-\ov U^l)+
\h\ \lf[\ \psi(1-\th^l_{ab})+\psi(\th_{ab}^l)\ \ri]\ri)\ri\},\ i\neq j\ ,}}
with $(a,b)=(1,2)$ or $(1,3)$ and where $\psi_n$ is the $n$'th Polygamma function, 
with $\psi_0\equiv \psi$.
Obviously, $\partial_M\theta^1_{12}=\partial_M\theta^1_{13}=0$, and
\eqn\dthetaab{\eqalign{
\partial_M\theta^2_{12}&=\partial_M\theta^2_{13}={1\over
\pi}{\alpha'g\over {(\alpha'g)^2+({\rm Im}{\cal T}^2)^2}}\ \partial_M({\rm
Im}{\cal T}^2)\ ,\cr
\partial_M\theta^3_{12}&=\partial_M\theta^3_{13}=-{1\over
\pi}{\alpha'g\over {(\alpha'g)^2+({\rm Im}{\cal T}^3)^2}}\ \partial_M({\rm
Im}{\cal T}^3)\ ,\cr
\partial_M\partial_{\ov N}\theta^2_{12}&=\partial_M\partial_{\ov N}\theta^2_{13}={1\over
\pi}{\alpha'g\over [(\alpha'g)^2+({\rm Im}{\cal T}^2)^2]^2}\ [\ -2\ ({\rm Im}{\cal T}^2)\ 
\p_M({\rm Im}{\cal T}^2)\ \p_{\ov N}({\rm Im}{\cal T}^2)\cr
&+\{(\alpha'g)^2+({\rm Im}{\cal T}^2)^2\}\ \p_M\p_{\ov N}({\rm Im}{\cal T}^2)\ ]\ ,\cr
\partial_M\partial_{\ov N}\theta^3_{12}&=\partial_M\partial_{\ov N}\theta^3_{13}=-{1\over
\pi}{\alpha'g\over [(\alpha'g)^2+({\rm Im}{\cal T}^3)^2]^2}\ [\ -2\ ({\rm Im}{\cal T}^3)\ 
\p_M({\rm Im}{\cal T}^3)\ \p_{\ov N}({\rm Im}{\cal T}^3)\cr
&+\{(\alpha'g)^2+({\rm Im}{\cal T}^3)^2\}\ \p_M\p_{\ov N}({\rm Im}{\cal T}^3)\ ]\ .}}
After substituting equations \deriv, \doublederiv\ and \dthetaab\ back into \twistedseven, we 
find after some algebra:
\eqn\twistedsevenb{\eqalign{
R^{7_1,7_2}_{M\ov N}&={G_{C^{7_17_2}\ov C^{7_17_2}}\over\im M\ \im N}\ \lf[X_{M\ov N}+\sum_{i=2,3}
\fc{g\ap}{\pi\ [(g\alpha')^2+(\im\Tc^i)^2]^2}\ri.\cr
&\times \lf\{s_i\ \lf(\gamma\ \ln(T^i-\ov T^i)-\ln(U^i-\ov U^i)+
\h\ [\ \psi_0(\theta^i_{ab})+\psi_0(1-\theta^i_{ab})]\ \ri)\ \ri.\cr
&\hskip1cm\times 
\lf(\ [\beta_i(M,N)-2\ \alpha_i(M,N)]\ (\im\Tc^i)^3+\beta_i(M,N)\ (\alpha'g)^2\ \im\Tc^i\ \ri)\cr
&\lf.\lf.+\h\ {\alpha'g\over \pi}\ [\ \psi_1(\theta^i_{ab})-\psi_1(1-\theta^i_{ab})\ ]\ 
 \alpha_i(M,N)\ (\im\Tc^i)^2\ \ri\}\ri]\ .}}
Here, the $\psi_n$ are the $n$'th Polygamma functions. In addition, we have introduced
the factor $s_i$, with $s_1=0,s_2=1,\ s_3=-1$ and:
\eqn\havedefined{\eqalign{
X_{S\ov S}&=\fc{1}{16}\ (1-6\beta-4\gamma)\ ,\cr
X_{S\ov T^j}&=-\fc{s_j}{8\pi}\ \gamma\ \fc{\alpha'g}{(\alpha'g)^2+({\rm Im}{\cal T}^j)^2}\ 
\im\Tc^j,\cr
X_{T^j\ov T^j}&=\fc{1}{16}\ \lf[1+2\beta+4\gamma\ (1-\th^j_{ab})\ri]-
\fc{s_j}{4\pi}\ \gamma\ \fc{\alpha'g}{(\alpha'g)^2+({\rm Im}{\cal T}^j)^2}\ \im\Tc^j\ ,\cr
X_{T^i\ov T^j}&=\gamma\ \fc{\ap g}{8\pi}\ \lf[s_i\ \fc{\im\Tc^i}{(\alpha'g)^2+({\rm Im}{\cal T}^i)^2}+
s_j\ \fc{\im\Tc^j}{(\alpha'g)^2+({\rm Im}{\cal T}^j)^2}\ri]\ \ ,\ \ i\neq j\ .}}

For the $1/2\ BPS$ sector $(23)$, the calculation is simpler and leads to:
\eqn\halfRie{
R^{7_27_3}_{M\ov M}=-\h\ {1\over (M-\ov M)^2}\ G_{C^{7_27_3}\ov C^{7_27_3}}
}
for $M=S,\,T^1,\,U^2,\,U^3$, for the other moduli, the components are zero.

Finally, Table A.1 shows the quantities
$$\beta_j(M,N)=\fc{\im M \ \im N}{\im\Tc^j}\ \fc{\p^2\im\Tc^j}{\p M\p \ov N}\ \ \ ,\ \ \ 
\alpha_j(M,N)=\fc{\im M \ \im N}{(\im\Tc^j)^2}\ \fc{\p\im\Tc^j}{\p M}\ 
\fc{\p\im\Tc^j}{\p \ov N}\ ,$$
\vskip0.2cm
{\vbox{\ninepoint{$$
\vbox{\offinterlineskip\tabskip=0pt
\halign{\strut\vrule#
&~$#$~\hfil
&\vrule$#$
&~$#$~\hfil
&\vrule$#$
&~$#$~\hfil
&\vrule$#$
&~$#$~\hfil
&\vrule$#$
&~$#$~\hfil
&\vrule$#$\cr
\noalign{\hrule}
&\ (M,N)&&\ \beta_2&&\ \alpha_2&&\ \beta_3&&\ \alpha_3&\cr
\noalign{\hrule}\noalign{\hrule}
&(S,S)&&\ {3\over 16}  &&\ {1\over 16}&&\  {3\over 16}&&\  {1\over 16}&\cr
&(S,T^1)    &&\  {-1\over 16}  &&\  {-1\over 16}&&\  {-1\over 16}&&\  {-1\over 16}&\cr
&(S,T^2)&&\ {1\over 16}  &&\ {1\over 16}&&\ {-1\over 16}&&\ {-1\over 16}&\cr
&(S,T^3)     &&\ {-1\over 16}  &&\ {-1\over 16}&&\ {1\over 16}&&\ {1\over 16}&\cr
&(T^1,T^1)     &&\ {-1\over 16}  &&\ {1\over 16}&&\ {-1\over 16}&&\ {1\over 16}&\cr
&(T^1,T^2)    &&\ {-1\over 16}   &&\ {-1\over 16}&&\ {1\over 16}&&\ {1\over 16}&\cr
&(T^1,T^3) &&\ {1\over 16}  &&\ {1\over 16}&&\ {-1\over 16}&&\ {-1\over 16}&\cr
&(T^2,T^2)&&\ {3\over 16}  &&\ {1\over 16}&&\ {-1\over 16}&&\ {1\over 16}&\cr
&(T^2,T^3) &&\ {-1\over 16}  &&\ {-1\over 16}&&\ {-1\over 16}&&\ {-1\over 16}&\cr 
&(T^3,T^3)&&\ {-1\over 16}  &&\ {1\over 16}&&\ {3\over 16}&&\ {1\over 16}&\cr
\noalign{\hrule}}}$$
\vskip-6pt
\centerline{\noindent{\bf Table A.1}
{\ }}
\centerline{}
\vskip10pt}}}
\vskip-0.5cm \ \br
while Table A.2 displays
$$\gamma_j(M)=\fc{\im M}{\im\Tc^j}\ \fc{\p\im\Tc^j}{\p M}$$
\vskip0.2cm
{\vbox{\ninepoint{$$
\vbox{\offinterlineskip\tabskip=0pt
\halign{\strut\vrule#
&~$#$~\hfil
&\vrule$#$
&~$#$~\hfil
&\vrule$#$
&~$#$~\hfil
&\vrule$#$\cr
\noalign{\hrule}
&\ M &&\ \gamma_2&&\ \gamma_3&\cr
\noalign{\hrule}\noalign{\hrule}
&S&&\ {i\over 4}  &&\ {i\over 4}&\cr
&T^1    &&\ {-i\over 4}  &&\ {-i\over 4}&\cr
&T^2&&\ {i\over 4}  &&\ {-i\over 4}&\cr
&T^3     &&\ {-i\over 4}  &&\ {i\over 4}&\cr
\noalign{\hrule}}}$$
\vskip-6pt
\centerline{\noindent{\bf Table A.2}
{\ }}
\centerline{}
\vskip10pt}}}
\vskip-0.5cm \ \br
 
\listrefs
\end